\documentclass[useAMS,usenatbib,referee]{mn2e}
\usepackage{lscape}
\usepackage{deluxetable}
\usepackage{lastpage}

\def\ms{ms$^{-1}$}

\def\mj{M$_{\rm{J}}$}

\title[Improved Periodogram Techniques]{Improved signal detection algorithms for unevenly sampled data.  Six signals in the radial velocity data for GJ876}
\author[J.S. Jenkins et al.]{J.S. Jenkins$^1$\thanks{E-mail: jjenkins@das.uchile.cl}, N. Becerra Yoma$^2$, P. Rojo$^1$, R Mahu$^2$, J. Wuth$^2$ \\
$^1$Departamento de Astronomia, Universidad de Chile, Camino el Observatorio 1515, Las Condes, Santiago, Chile, Casilla 36-D\\
$^2$Dept. of Electrical Engineering, Universidad de Chile, Av. Tupper 2007, PO Box 412-3, Santiago, Chile}
\begin{document}

\date{Submitted January 2011}

\pagerange{\pageref{firstpage}--\pageref{LastPage}} \pubyear{2002}

\maketitle

\label{firstpage}

\begin{abstract}

The hunt for Earth analogue planets orbiting Sun-like stars has forced the introduction of novel methods to 
detect signals at, or below, the level of the intrinsic noise of the observations.  We present a new global periodogram method 
that returns more information than the classic Lomb-Scargle periodogram method for radial velocity signal detection.  Our method 
uses the Minimum Mean Squared Error as a framework to determine the optimal number of genuine signals present 
in a radial velocity timeseries using a global search algorithm, meaning we 
can discard noise spikes from the data before follow-up analysis.  This method also allows us to determine the phase 
and amplitude of the signals we detect, meaning we can track these quantities as a function of time to test if the 
signals are stationary or non-stationary.  
We apply our method to the radial velocity data for GJ876 as a test system to highlight how the phase 
information can be used to select against non-stationary sources of detected signals in radial velocity data, such as rotational modulation 
of star spots.  Analysis of this system yields two new statistically significant signals in the combined Keck and HARPS 
velocities with periods of 10 and 15~days.  Although a planet with a period of 15~days would relate to a Laplace resonant chain 
configuration with three of the other planets (8:4:2:1), we stress that follow-up dynamical analyses are needed to test 
the reliability of such a six planet system.

\end{abstract}


\section[intro]{Introduction}

The discovery and characterisation of low-mass rocky planets is moving at a fast pace, now that 
technology has reached the limit necessary to discover these elusive objects.  Along with the 
technological advances, a number of stars have been monitored with long enough time baselines to gain the necessary 
clutch of data points that allows low-amplitude signals to be announced with a high level of 
statistical significance.

Teams discovering these planetary systems generally use the classic method of applying 
a Lomb-Scargle periodogram (LSP) algorithm to their data, which looks for the presence of sine and 
cosine functions buried in any discretely sampled timeseries (\citealp{scargle82}).  This method has 
been very successful in finding low-amplitude signals that betray the presence of small planets 
(e.g. \citealp{mayor09}; \citealp{vogt10}), however it has recently been shown to be 
an inefficient way to detect such signals (\citealp{tuomi13a}).

At this point it is worth introducing a short explanation of where in the procedure of planet detection 
the LSP fits in.  Once the observations of a given target star have been converted into 
Doppler velocity points, the timeseries is fed to the LSP.  The periodogram performs a 
Fourier-like analysis of the data, and the observer then searches for the most powerful frequencies 
output by the algorithm.  The observer can then move to the next step, which is characterisation of any 
frequency found by the algorithm, proving if this frequency is a genuine Doppler shift or not.  Therefore, 
the LSP algorithm plays a fundamental role in planet detection, since if the power of any Doppler 
frequency is not highlighted, the planetary signal will remain hidden.

Another method that has recently been employed to search for low-amplitude signals in radial 
velocity data is through Bayesian analysis (\citealp{tuomi12}; \citealp{tuomi13a}), and this has 
given rise to a number of new, and low-mass, planetary systems (e.g. \citealp{anglada-escude12}; 
\citealp{tuomi13b}; \citealp{jenkins13a}).  Although, Bayesian analysis seems to detect low-amplitude 
signals more efficiently than application of the LSP method, it is not as easily applicable as the LSP method 
and is less intuitive, so is not widely used in the exoplanet community.

With this in mind, we have begun a program to move beyond the LSP, such that we can detect low-amplitude 
signals with greater efficiency, and with greater certainty than is currently possible.  In $\S$~2 and $\S$~3 we describe the 
LSP methodology, and our first step along the path to implementing better detection methods that surpass the LSP algorithm.  
In $\S$~4 we describe the comparison tests we did with the LSP, and highlight the 
areas where our algorithm can provide greater benefits.  In $\S$~5 we apply our algorithm to the 
radial velocity data set from Keck and HARPS for the star GJ876, which is known to host a dynamically 
stable system of at least four planets, three of which are in a Laplace resonance.  Finally, we summerise our code and results 
in $\S$~6.

\section[ls]{Lomb-Scargle Periodogram}

Classically in the exoplanet field, Keplerian orbits have been detected in radial velocity data by application 
of the LSP (\citealp{scargle82}).  The LSP was developed as a tool to statistically evaluate the level of 
significance of any detected periodic signal in unevenly sampled data by translating the data into the 
Fourier domain and searching for sine and cosine patterns that repeat.  Sine and cosine functions that are stable and well sampled manifest as strong 
peaks in the power spectrum of such data, and an observer can then select the strong peaks that pass some 
significance threshold, and study the velocities for the presence of planets.  The complications arise from the 
spectral window function and the additive noise in the data.

Although the LSP has been an extremely successful tool in astrophysics, the method has some drawbacks, which 
we will explain here in the context of the search for exoplanets.  
The LSP base algorithm is shown in Eq$^{n}$~\ref{eq:lsp} and one can see that the method performs a Fourier transform of 
the data and then searches for periodic sinusoidal frequencies using a sine and cosine based method:

\begin{equation}
\label{eq:lsp}
P_{X}(\omega) = \frac{1}{2} \left\{\frac{\Big\lbrack  \sum \limits_{j} X_{j} ~cos ~\omega (t_{j}-\tau) \Big\rbrack^{2}}{ \sum \limits_{j} cos^{2} \omega (t_{j}-\tau)}
+ \frac{\Big\lbrack  \sum \limits_{j} X_{j} ~sin ~\omega (t_{j}-\tau) \Big\rbrack^{2}}{ \sum \limits_{j} sin^{2} \omega (t_{j}-\tau)}\right\}
\end{equation}

where $P_{X}$ is the periodogram powers as a function of frequency, $\omega$ is the sinusoidal frequencies, $t_{j}$ are the timestamps of 
the observations, and $\tau$ is defined as follows:

\begin{equation}
\label{eq:lsp1}
tan(2\omega\tau) = \frac{ \left(  \sum \limits_{j} sin 2 \omega t_{j} \right) }{\left(  \sum \limits_{j} cos 2 \omega t_{j} \right) }
\end{equation}

The power of this algorithm comes from the fact that when the data ($X$) contains a sinusoidal component 
that has a characteristic frequency $\omega_{0}$, then around $\omega = \omega_{0}$, this frequency is in phase 
with the introduced sinusoidal factors in the Fourier domain, and the signal strength is increased.  The signal then 
appears as a strong peak in the power domain when compared to other peaks across the full frequency window that 
has been sampled.

One issue with this method is that the sine and cosine functions applied in the signal search are not strictly orthogonal for unevenly sampled 
data, meaning that the power spectral points, $P_{X}$, are not independent variables, which means they are correlated.  \citet{scargle82} 
discussed this issue and was able to show that this dependence can be limited by construction of a well chosen frequency grid to sample the 
Fourier domain.  However, when applied to low S/N data, such as low-amplitude signals in radial velocity data produced by orbiting low-mass 
planets, this inter-dependency in the power spectrum complicates the signal detections, particularly for a superposition of low-amplitude 
signals arising from multiple low-mass planets.

Another issue in the search for planetary orbits from radial velocity data is that there are a high fraction of reported planets that are not in 
circular orbits that are well described by single sine and cosine functions, but rather require a more complex Keplerian function to describe them.  
\citet{otoole09} introduced a new method to search for frequencies in radial velocity data that directly employs a Keplerian search algorithm, 
following a number of the descriptions in \citet{cumming04}, that appears to be significantly better at detecting signals from planets on 
highly eccentric orbits.

Additionally, in the LSP the noise properties for the data $X$ are assumed to be white when computing the statistical significance of each 
sampled frequency point.  For precision radial velocity work, it is now known that the velocities themselves 
can exhibit significant correlations (see \citealp{baluev13}; \citealp{tuomi13a}), correlations that should be accounted for.  
Therefore, in the strictest sense, analytical FAPs calculated using the equations laid out in \citet{scargle82} are incorrect, given the 
underlying assumption of Gaussian distributed noise.  Therefore, the presence of correlated noise increases the difficulty of detecting 
signals using an approach that does not consider that noise in any way.

More practically, given the method of signal detection in exoplanetary science, the normal multi-planet detection methods introduce 
problems.  The standard approach is as follows:

\begin{enumerate}
 \item Apply the LSP to the radial velocities
  \item Locate the strongest and significant signal at a given frequency (orbital period)
  \item Fit a Keplerian function to the velocities around that period and subtract off to get the residuals
  \item Re-apply the LSP to the residuals to search for additional signals in the data
   \item Repeat until the white noise floor is reached or there are no significant peaks remaining
\end{enumerate}

This procedure is not optimal for detecting low-amplitude signals buried in the noise of radial velocity timeseries data sets because 
once a signal has been detected and fit out, a search for additional signals in the residuals will depend on the original model that was subtracted 
off, since generally these signals are not orthogonal.  New approaches have been 
introduced to try to circumvent this issue, such as applying Bayesian inference, mentioned above, or recursive periodogram methods 
(\citealp{baluev13}; \citealp{anglada-escude12b}).

Finally, the LSP by definition is a method that works by measuring the harmonic content of any signals present in the power spectrum 
of a data set, which means that there is no information provided by the algorithm on the amplitude or phase of the signal found in 
the data.  This means that any signal detected by the LSP that is then searched for by using a least-squares Keplerian fitting procedure 
in the velocities, is not gaurenteed to be the same signal.  In real terms it is highly likely that the signal fit is the same signal, but without 
additional information from the LSP, it is difficult, if not impossible, to prove this.  In addition, the amplitude and phase of a signal contains vital 
information, since they can be used to show that any signal detected is a true \emph{stationary} Doppler signal, or is a quasi-stationary signal 
that could arise from time-dependent phenomena on the star such as spot rotation (e.g. \citealp{dawson10}).

Given the limits of the LSP method for low-amplitude signal detection in unevenly sampled data mentioned above, we have 
devised a method based around the minimum mean square error (MMSE) that can allow 
the period, amplitude, phase, and number of components that best describes the data set, to be found directly from the search algorithm 
itself.  The following section is devoted to explaining our methodology.  We note that \citet{zechmeister09} introduced the Generalised 
Lomb Scargle Periodogram method that handles better more eccentric Keplerian orbits.  However, our MMSE method has specifically been 
designed to determine the number of real components in a data set in a global fashion, in order to aid in the detection of the emerging 
population of low-mass and circular multi-planet systems.  However, it is fairly trivial to add an extra offset component to the following 
MMSE algorithm that will perform the same function as the Generalised LSP.

\section[algo]{Minimum Mean Square Error}

Our MMSE method utilises five steps sequentially in order to estimate the most significant components 
in a sequence of samples generated with non-uniform sampling.  Firstly, a Fourier-like analysis is run 
on the data, using the MMSE, and the components with the lowest square error are selected from 
the resultant MMSE series.  Secondly, the local neighbourhoods of those components are selected for 
scrutiny.  Next, the MMSE is then recomputed for all possible combinations of components and their 
neighbours, selected in the previous steps.  Finally, the combination with the lowest MMSE is chosen as 
the most significant sinusoidal components of the given data sets.

We introduce our analysis method using the MMSE in the following manner.  Given a set of data, such as 
radial velocity measurements that contain any number of frequencies that we will assign $x(i)$, where 
1$\le i \le I$ and $I$ is the length of the data set, we can estimate the number of sinusoidal components 
$N_{C}$, and the set $S = \left\{ C_1,C_2,...,C_i,...,C_{N_{C}} \right\}$ where $C_i$ corresponds to the $i$-th 
sinusoidal component, by formulating the set of frequencies like Eq$^{\rm{n}}$~\ref{eq:set}:

\begin{equation}
\label{eq:set}
S = \left\{(\omega_1,a_1,\phi_1),...,(\omega_i,a_i,\phi_i),...,(\omega_{N_{C}},a_{N_{C}},\phi_{N_{C}}) \right\}
\end{equation}

where ($\omega_i$,$a_i$,$\phi_i$) corresponds to the frequency, amplitude, and phase of the $i$-th component 
in $C_i$.  Therefore, by application of the MMSE method, the problem can be formulated as finding the set 
$S = \left\{ (\omega_i,a_i,\phi_i) \right\}^{N_{C}}_{i=1}$ that satisfies the following minimisation:

\begin{equation}
\label{eq:mmse}
S = \arg\!\min_{\omega_i,a_i,\phi_i}\left\{ \sum\limits_{t=1}^T \left( X(t) - \sum\limits_{i=1}^{N_{C}} a_i cos(\omega_i t + \phi_i) \right)^2  \right\}
\end{equation}

where the sinusoidal components present in $X(t)$ are given by the set of triplets with the minimum square 
error, and $N_{C}$ corresponds to the number of triplets present in the analysis.

Solving Eq.~\ref{eq:mmse} requires an exhaustive search over $N_{C}$, $\omega$, $a$, and $\phi$.  Even if $N_{C}$ were 
known a priori, the number of possible combinations of $\omega$, $a$, and $\phi$ are beyond computational 
capacity.  Let the target frequency bandwidth, the amplitude, and the phase be divided into $K_{\omega}$, 
$K_{a}$, and $K_{\phi}$ levels respectively, then the number of possibilities for taking $N_{C}$ triplets 
($\omega$, $a$, $\phi$) simultaneously is ($K_{\omega} \cdot K_{a} \cdot K_{\phi}$)$^{N_{C}}$.  Since all 
possible values of $\omega$, $a$, and $\phi$ have to be explored, the program is unfeasible from the 
point of view of computational load.  For instance, if we consider high precision in frequency ($K_{\omega} 
\approx 10^3$) and moderate precision in amplitude and phase ($K_a \approx 10^2, K_{\phi} \approx 10^2$), 
the number of possible combinations of $\omega$, $a$, and $\phi$ grow rapidly, since for $N_{C}$ = 7, and considering 
that each combination can be evaluated in one CPU instruction, 10$^{49}$ instructions are necessary, 
which would require more than 10$^{39}$ seconds or approximately 3$\times$10$^{31}$ years on a 10~GHz 
processor, clearly not a calculation that can be done in any reasonable time frame.  As an alternative, 
a MMSE based method is proposed since it can be easily implemented in multi-core computing, reducing 
the processing time drastically.

\underline{\bf Step 1:} The first step is to perform a MMSE based Fourier analysis, where the target frequency bandwidth is divided 
into $K_{\omega}$ levels.  Each level $\omega_{k}$ is represented by 

\begin{equation}
\label{eq:omega}
\omega_{k} = \frac{\pi * k}{K_{\omega}}, ~~~~~~{\rm where} ~~1~\le~k~\le~K_{\omega}
\end{equation}

For each $\omega_{k}$ an optimal amplitude $a_{\omega_{k}}$ and phase $\phi_{\omega_{k}}$ are computed 
according to the following equation:

\begin{equation}
\label{eq:omega1}
(a_{\omega_{k}},\phi_{\omega_{k}}) = \arg\!\min_{a,\phi} \left\{ \sum^{T}_{t=1} \left( X(t) - a cos(\omega t + \phi) \right)^2  \right\}
\end{equation}

By application of Eq$^{n}$~\ref{eq:omega1}, a set $S_{P}$ of components is obtained as shown in 
Eq$^{n}$~\ref{eq:speri}

\begin{equation}
\label{eq:speri}
S_{P} = \left\{ (\omega_{k}, a_{\omega_{k}}, \phi_{\omega_{k}}) \right\}^{K_{\omega}}_{k=1}
\end{equation}

This set corresponds to the new periodogram of the data.

\underline{\bf Step 2:} Now one must select the $N$ components that show the lowest MMSE, either by visual 
inspection or by automatically selecting all MMSE values below a threshold 
level selected at the discretion of the user, or based on some noise threshold criteria (e.g. \citealp{kuschnig97}).  
From the set $S_{P}$ obtained in the previous step, a subset 
$S_{min} = \left\{ (\omega_{i}, a_{\omega_{i}}, \phi_{\omega_{i}}) \right\}$ is built with the $N$ components 
that show the lowest MMSE or the highest amplitudes.

\underline{\bf Step 3:} For each component $C_{i}$ associated with the triplet ($\omega_i$, $a_{\omega}$, $\phi_{\omega}$) 
in $S_{min}$, a neighbourhood $V_i$ is defined, according to Eq$^{\rm{n}}$~\ref{eq:neigh}

\begin{equation}
\label{eq:neigh}
V_i = \left\{  (\omega, a_{\omega}, \phi_{\omega}) \in S_{P} / \omega \in [\omega_i - \delta, \omega_i + \delta]  \right\}
\end{equation}

where $\delta$ is defined as the number of elements of $V_i$ that are lower than $N_{\rm{max}}$, and $N_{\rm{max}}$ 
defines the maximum number of elements of any neighbourhood.  Therefore, $\delta$ is set to some value 
to incorporate all the significant $S_{P}$ peaks.  As a result, each component 
$C_i(\omega_i, a_{\omega}, \phi_{\omega})$ provides $M_{C_i}$ candidates, where $M_{C_i}$ is given by the 
cardinality of $V_i$.

We note that an alternative to the neighbourhood defined around each component $C_{i}$, the trellis analysis can be performed by taking into 
consideration the frequency of $C_{i}$ only.  In this case, another neighbourhood can be defined by considering sets of amplitudes and 
phases to assess the validity of $C_{i}$ as a true component of the data set.

\begin{figure}
\vspace{6.5cm}
\hspace{-4.0cm}
\includegraphics{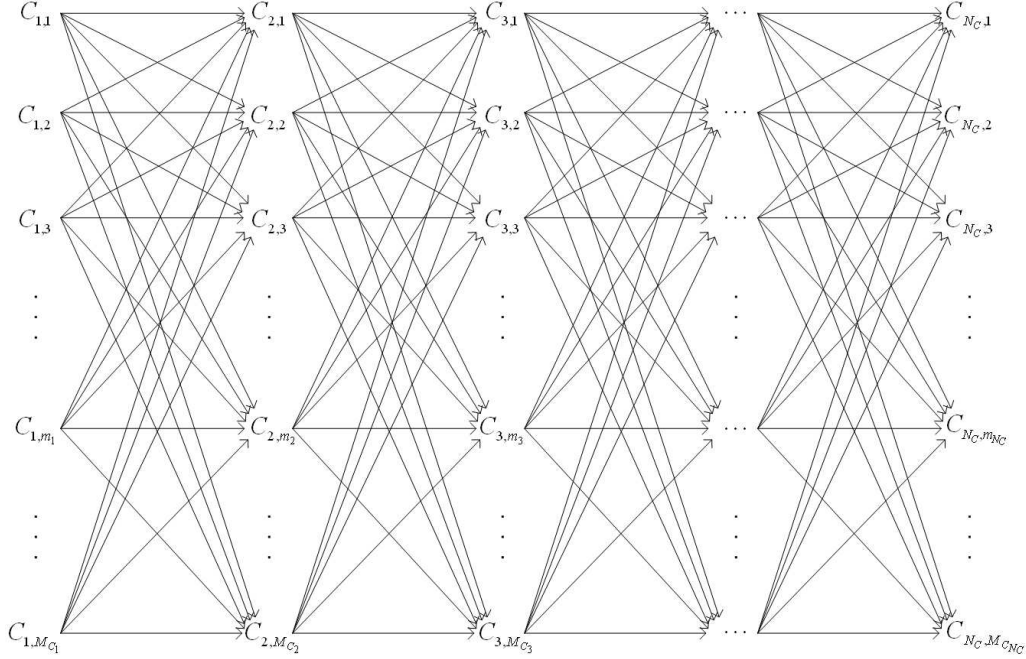}
\vspace{2cm}
 \caption{An example trellis diagram, illustrating all the possible combinations of components and their local neighbourhoods.}
\label{trellis}
\end{figure}

\underline{\bf Step 4:} The MMSE is computed for all the possible combinations of $N_C$ components, where 
$N_{C}$ is lower than or equal to $N$.  A trellis is generated with all the candidates per component (see 
Fig.~\ref{trellis}).  For each $N_C < N$, let $A^j = \left\{  C^j_1, C^j_2, ..., C^j_{N_{C}} \right\}$ be a set 
with one of the combinations of $N_C$ components from $N$, where 1 $< j < \frac{N!}{N_C!(N - N_C)!}$ and $C^j_i$ corresponds 
to the $i$th component in the $j$th combination.  The corresponding set of neighbourhoods is $V_{A^j} = \left\{  V^j_1, V^j_2, ..., V^j_{N_C}  \right\}$ 
where $V^j_i$ denotes the neighbourhood of components $C^j_i$.  For 
each $A^j$, the set of $N_C$ triplets ($\hat{\omega}^j_1, \hat{a}^j_1, \hat{\phi}^j_1$), ...., 
($\hat{\omega}^j_{N_C}, \hat{a}^j_{N_C}, \hat{\phi}^j_{N_C}$) is found whereby the lowest MMSE is obtained according to 
Eq$^{\rm{n}}$~\ref{eq:mmse_neigh}.

\begin{equation}
\label{eq:mmse_neigh}
(\hat{\omega}^j_1, \hat{a}^j_1, \hat{\phi}^j_1), ...., (\hat{\omega}^j_{N_C}, \hat{a}^j_{N_C}, \hat{\phi}^j_{N_C}) = \arg\!\min_{(\omega^j_i, a^j_i, \phi^j_i) 1~\le~i~\le~N_C} \sum^T_{t=1} 
\left( X(t) - \sum^{N_C}_{t=1} a^j_i cos(\omega^j_i t + \phi^j_i) \right)^2
\end{equation}

where ($\hat{\omega}^j_i, \hat{a}^j_i, \hat{\phi}^j_i$) $\in V^j_i$, 1~$\le$~i~$\le~N_C$ and in this way, an optimal set of 
$N_C$ triplets is obtained for each $N_C$.

\underline{\bf Step 5:} Finally, one must choose the $N_C$ that minimises the MMSE.  To do this we let 
$S = \left\{ (\hat{\omega}_1, \hat{a}_1, \hat{\phi}_1), ..., (\hat{\omega}_{N_C}, \hat{a}_{N_C}, \hat{\phi}_{N_C}) \right\}$ 
be the set with the lowest MMSE obtained in the previous step.  The most important sinusoidal components 
present in the data set are given by the $N_C$ triplets in $S$.

\section[tests]{Functionality Tests}

In order to test the reliability of the algorithm we describe above, we ran two tests on simulated data 
where 1) we simulate a pure signal and 2) we simulate superposition of five signals without harmonics present.  
We also run the same tests on the LSP and compare the results from both.

\begin{figure*}
\vspace{4.5cm}
\hspace{-4.0cm}
\includegraphics{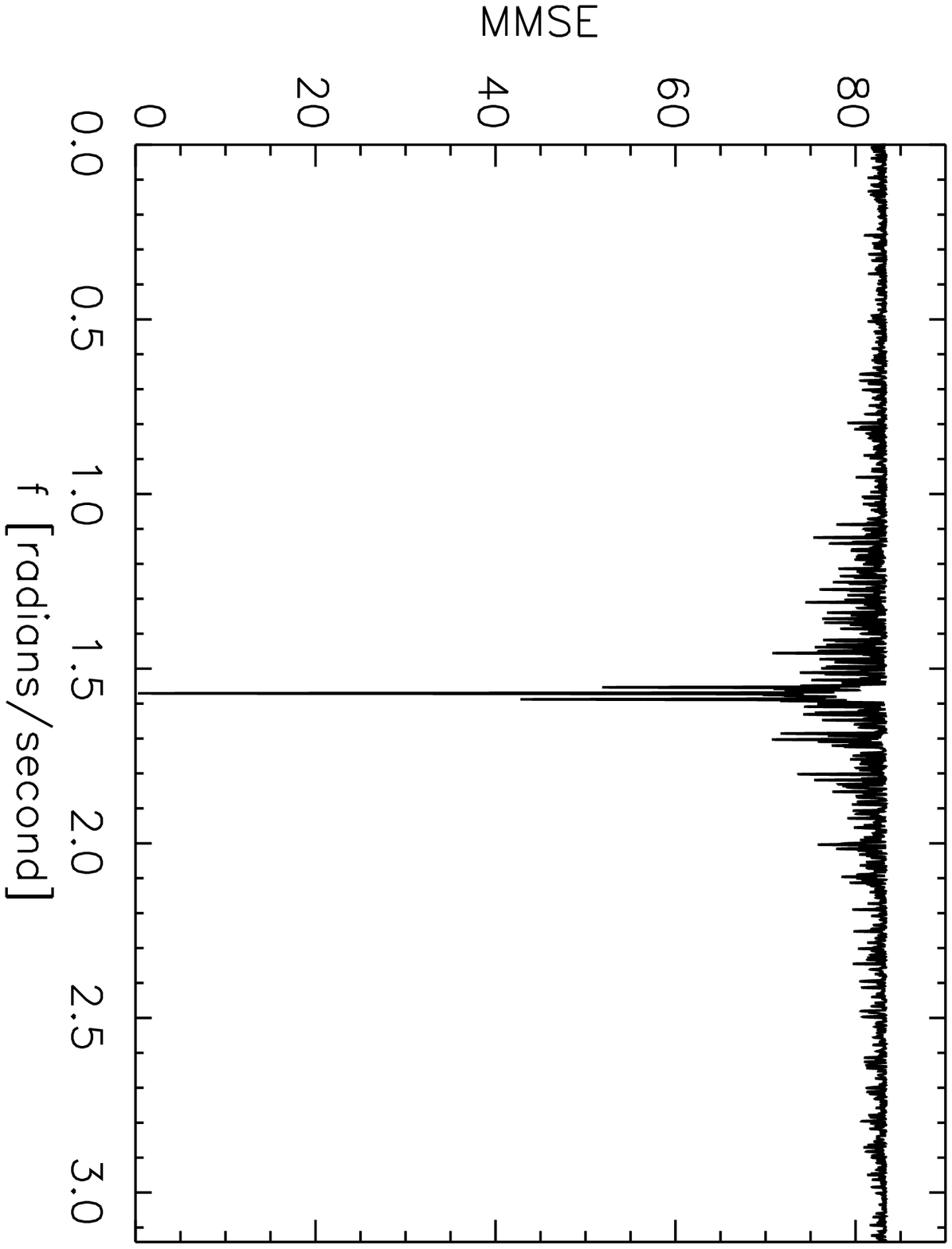}
\includegraphics{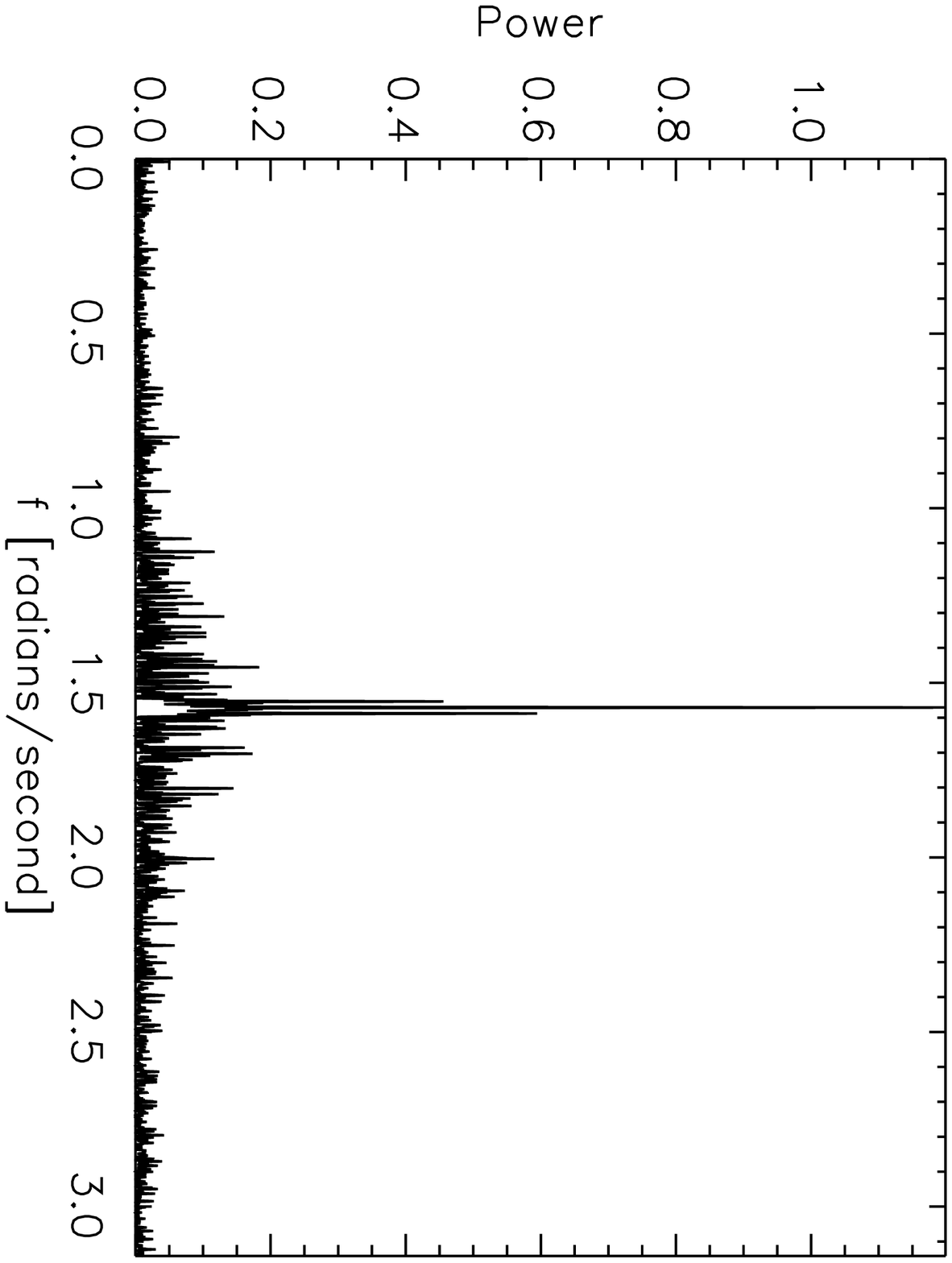}
\includegraphics{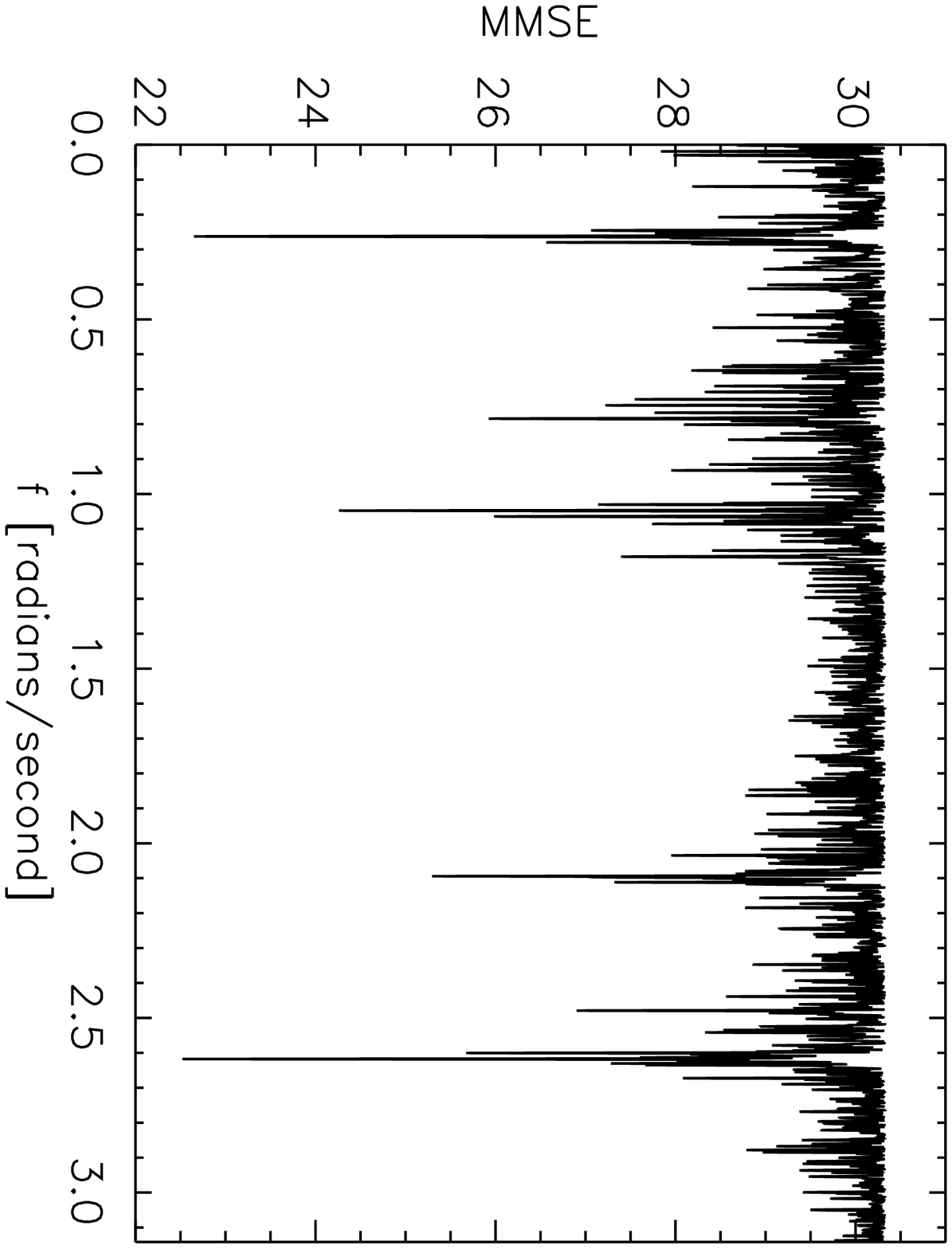}
\includegraphics{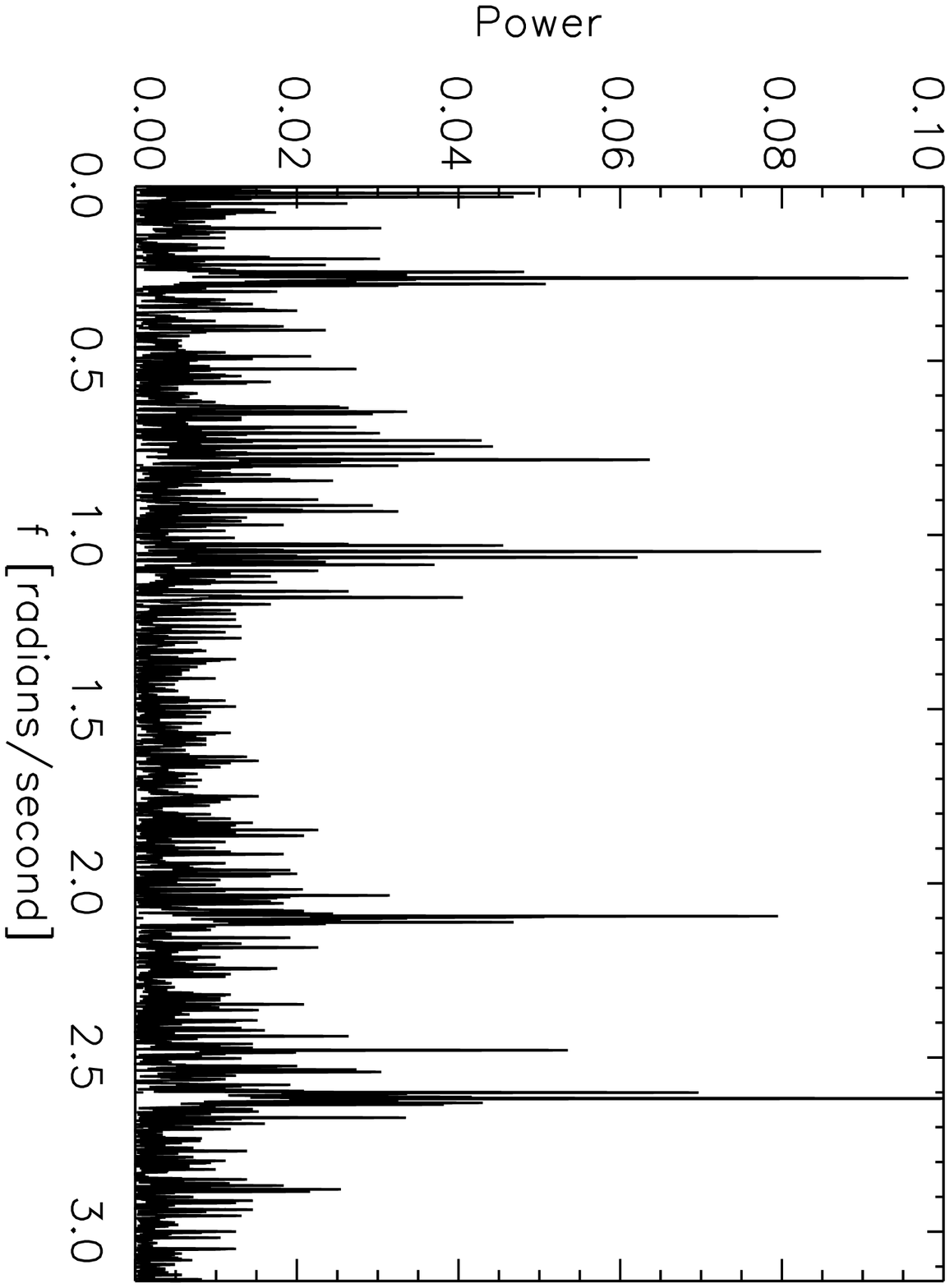}
\vspace{6.8cm}
\caption{The left column corresponds to the MMSE periodograms, where the detection of a simulated pure signal (top) 
and a signal with harmonics included (bottom) are shown.  The right column shows the same signals detected with the LSP.}
\label{signals}
\end{figure*}

\subsection{Pure Signal}

In the top panel of Fig.~\ref{signals} we compare the MMSE values against the LSP powers as a function of 
frequency for a pure signal implanted in a simulated radial velocity timeseries with uneven sampling.  The 
signal was generated simply as $a*cos(\omega t+\phi)$, where $a$=1.0998, $\omega= \pi/2$, and $\phi$=2.  
We note that with even sampling the signal is significantly stronger and the noise spikes are heavily suppressed as they 
primarily arise due to interference from the window function from unevenly sampled data.  

In both analyses we recover the signal with low false alarm probability (FAP) at a frequency of 0.25~Hz, or $\pi$/2 radians per second (1.5708~rads/sec), as shown in 
the top plots in Fig.~\ref{signals}.  The FAP's were calculated following a bootstrap method where we scrambled the velocities with replacement, computing the 
periodograms each time and comparing the strength of the strongest frequency to the observed frequency, recording the number of times this frequency was 
stronger than the observed frequency power.  This allowed us to calculate the probability that our observed strongest frequency could arise by chance given the data.  
The FAP's we calculated are both significantly $<$10$^{-4}$ and it can be seen that the noise falls off in a Lorentzian fashion when moving progressively further 
away from the central strong peak. 

\subsection{Tests without Harmonics}

The lower panels in Fig.~\ref{signals} show a similar simulated data set to the top panels, except five signals 
at various frequencies have been introduced without signal harmonics.  The signals 
were introduced as follows, $a_1*cos(\omega_1 t+\phi_1) + a_2*cos(\omega_2 t+\phi_2) + ... + a_5*cos(\omega_5 t+\phi_5)$, 
with parameter values of $a_1=a_2=...=a_5=0.3083$, $\phi_1=\phi_2=...=\phi_5=2$, and $\omega_i$ values, where 1$\le i \le$5, 
of $\pi/12, 3\pi/12, 4\pi/12, 8\pi/12$, and $10\pi/12$ radians, respectively.

It can be seen that the strength of these periodogram peaks are dramatically reduced, in comparison to the strength 
of the single pure peak found in the top panels.  In the LSP on the right, we 
see the five strong peaks across the frequency band of interest (0.0-0.5~Hz).  There are also secondary signals, 
not as strong as these five, but with significant power to decrease the significance of signal detection.  The 
introduced signals were recovered as the strongest peaks in this data, however after subtracting each one away, 
a process that is common in the exoplanet detection field, again the FAP's were extremely low, even in the presence of 
the other peaks.

The MMSE periodogram is very similar to the LSP, as expected.  The same patterns arise in the analysis, 
similar frequency peaks and surrounding noise peaks, and the FAP's of the signals are also similar to the LSP probabilities.  
Therefore, this shows that, to first-order, the MMSE is as powerful as the LSP for detecting multiple signals in a 
radial velocity timeseries, and the presence of additional signals in a data set does not adversely affect the MMSE 
method any more than it does the LSP.

\subsection{Component Test}

Above we have shown that the MMSE is as effective at detecting circular signals as the LSP, even though the 
MMSE also returns more information about the signals that have been tested.  We note that methods such as 
prewhitening can also return the period, amplitude, and phase of a signal, and prewhitening has been used 
to constrain the mass of Corot-7$b$ (\citealp{queloz09}; \citealp{hatzes10}) and to shed some doubt on the 
nature of the planet candidate orbiting $\alpha$~Centauri~B (\citealp{hatzes13}).  There are also packages 
available to perform this type of analysis, Period04 for example (\citealp{lenz04}).

\begin{figure*}
\vspace{4.5cm}
\hspace{-4.0cm}
\includegraphics{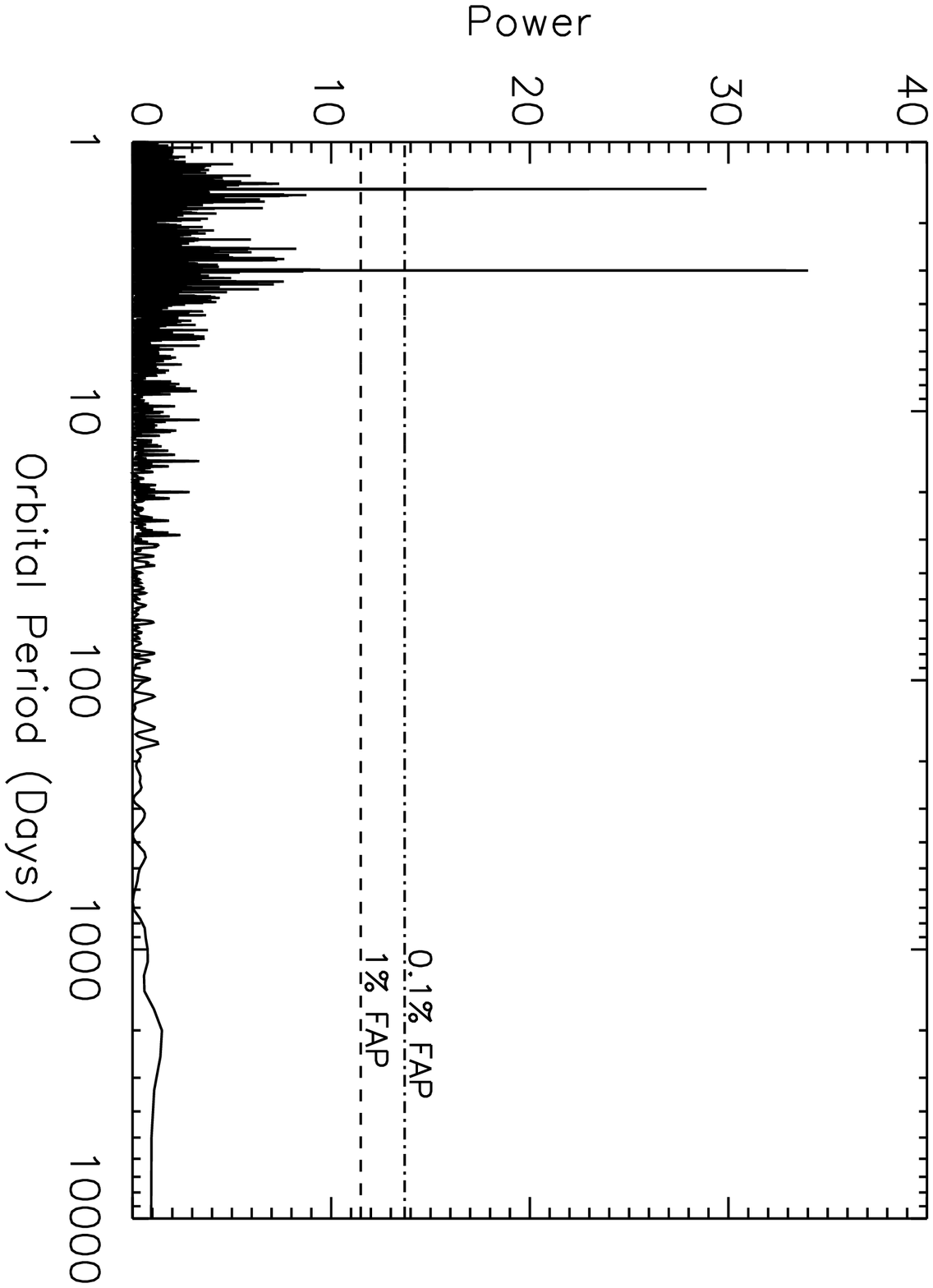}
\includegraphics{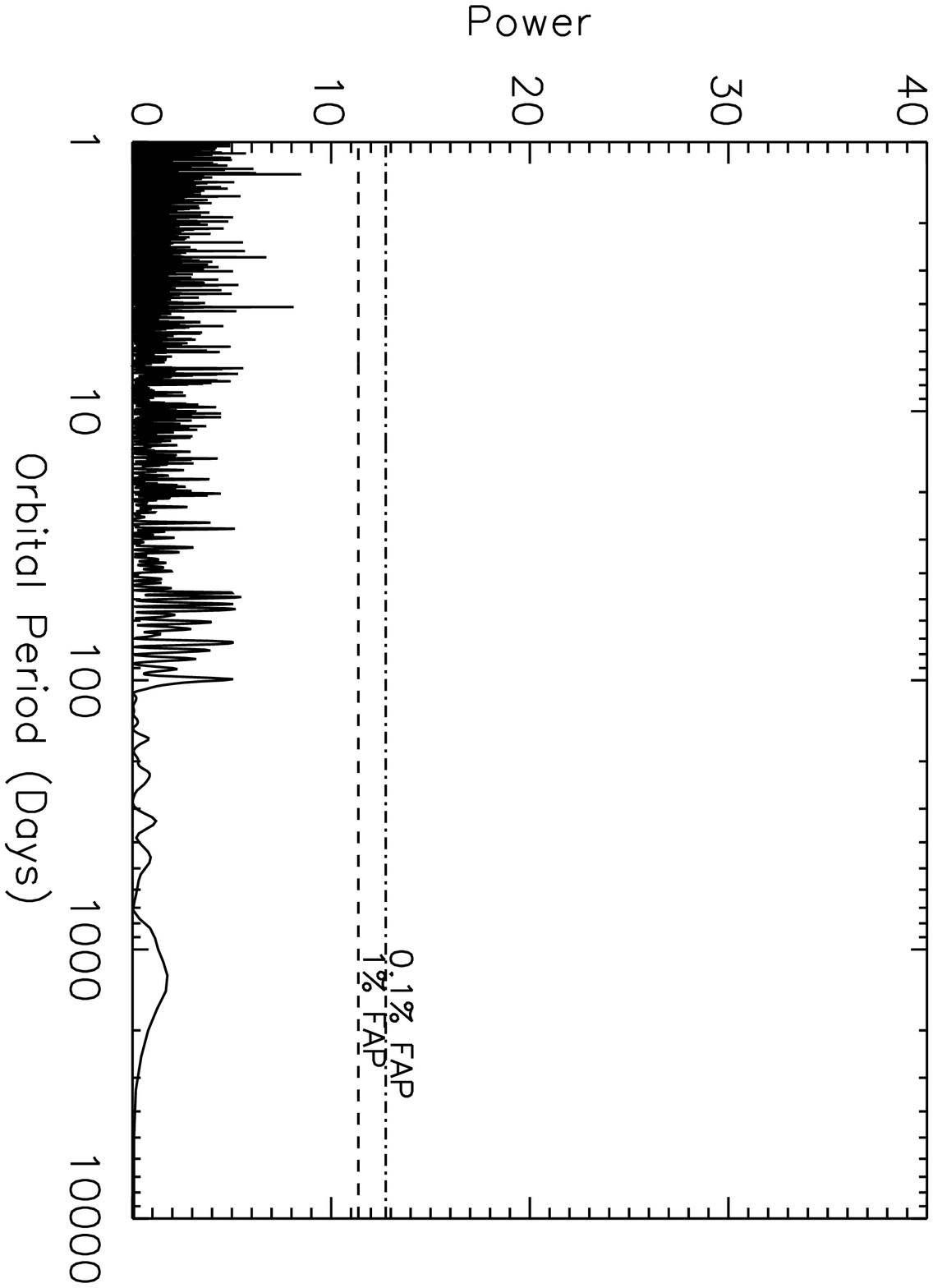}
\vspace{0.7cm}
\caption{Periodograms for a simulated system of two planets with periods of 3~days and 100~days and circular orbits.  
The left panel is the raw periodogram and right panel is the periodogram of the residuals after subtraction of the best fit 
orbit for the inner planet.  The FAP's for 1 and 0.1\% are shown for reference.}
\label{sim_peri}
\end{figure*}

As mentioned earlier, the downside to using methods such as prewhitening is that they are gradient based approaches that start 
by searching for one signal only, subtracting that signal out of the timeseries, and then searching the residuals 
all over again by treating the residuals as an independent timeseries from the original observed data, and then 
repeating this process until the noise floor is reached.  Our MMSE method on the other hand is 
a global approach that searches for all signals in a timeseries at the same time.  The value in this approach, 
at the cost of computing time, is that we make no assumption that the original timeseries only contains one 
signal, which is implicit in a gradient based approach, and hence we do not lose any of the power in secondary 
signals that is lost when subtracting out the primary frequency detected by the LSP/prewhitening approaches.

\begin{figure}
\vspace{4.5cm}
\hspace{-4.0cm}
\includegraphics{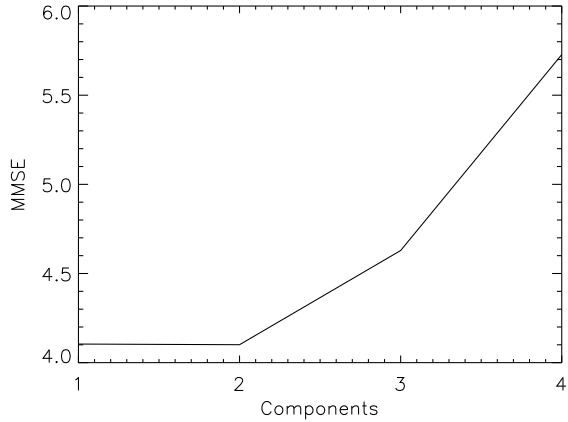}
\vspace{0.7cm}
\caption{Number of components detected by our trellis MMSE signal search in the two planet simulation.  The solid curve 
is minimised at two, centered on both signals in the simulated data set.}
\label{ncomp}
\end{figure}

As a test we simulated a system of two planets with orbital periods of 3 and 100 days, such that the high 
frequency signal was clearly detected in the original data but the lower frequency peak was not apparent, even after 
subtraction of the higher frequency signal (see Fig.~\ref{sim_peri}).  Following our MMSE approach, we then selected 
a number of the deepest spikes in the periodogram and performed the trellis component search part of our method.  
The search was able to determine that the data contained two, and only two, components, returning their periods, 
amplitudes, and phases.  The minimum of the curve in Fig.~\ref{ncomp} represents the detected number of signals in the 
data set.  This test shows that a global approach like this can be more powerful than a gradient based approach, 
particularly for detecting the emerging population of multi-planet low-mass systems.

\section{Application to the GJ876 Doppler Velocities}

GJ876 is a M-dwarf star that hosts a system of at least four planets, two of these planets are gas giants that have been found 
to be orbiting in a Laplace Resonance with one of the smaller planets in the system (\citealp{rivera10}).  Although, we expect that 
the resonance acts to rapidly alter the orbital elements of these planets, the two gas giants exert large velocity variations on the 
star large enough that both planets are easily detectable in the first and second halves of the data independently, and the signal 
is much larger than any variations in the elements we expect.  This represents a good test data set for our method as there are 
enough data points to detect multiple signals and the data sets come from two independent instruments and analysis methods.

\subsection{Reanalysis of Planetary System}

The planetary system around GJ876 has an important place in the history of exoplanet science, since it was the first planetary system 
shown to have planets in some resonant configuration (\citealp{marcy01}).  Resonances had been witnessed in the moons of Jupiter 
and various asteroid and planetary configurations in our Solar System.  It has 
been hypothesised that a previous 1:2 resonance of Jupiter and Saturn was the reason for the late heavy bombardment and the final 
positions of the outer giant planets (\citealp{tsiganis05}; \citealp{morbidelli05}).  However, the two gas giants orbiting GJ876 were the first exoplanets actually found to be in a 
resonant configuration, indicating that dynamical interactions of planetary bodies are indeed an important ingredient in the formation 
and evolution of planetary systems.

Currently four planets are known to orbit this star, with orbital periods of 1.94, 30.09, 61.12, and 124.26~days (\citealp{delfosse98}; \citealp{marcy01}; 
\citealp{rivera05}; \citealp{rivera10}), but there exists the exciting potential for more.  \citet{correia10} performed many 
dynamical simulations that suggest a planet with an orbital period of around 15~days could stably exist in a larger chain resonance, however 
they claimed that the radial velocity data did not support such a planet, and therefore if there is a planet with an orbital period of 15~days it 
can not have a mass any larger than around 2~M$_{\rm{\oplus}}$.  Such a result motivates the search for additional signals in the GJ876 
velocities.

\begin{figure}
\vspace{5.5cm}
\hspace{-4.0cm}
\includegraphics{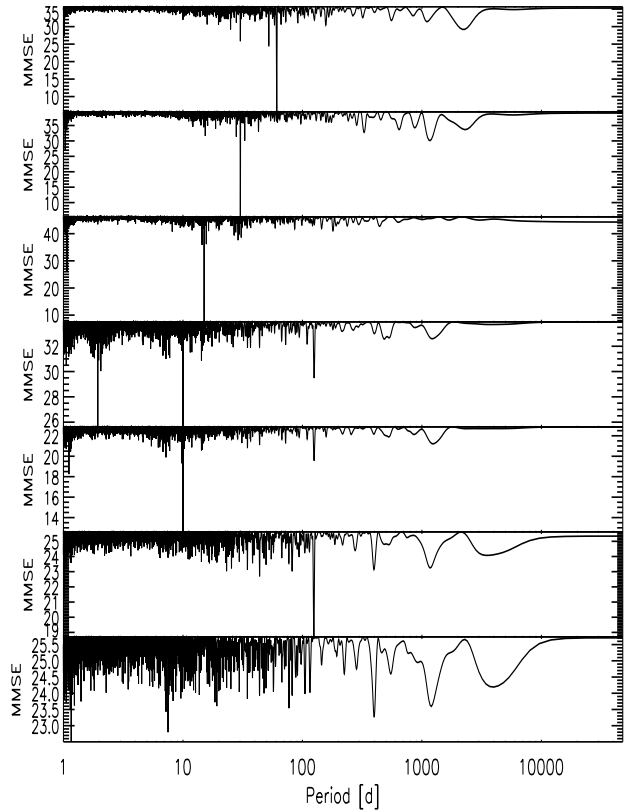}
\vspace{5.6cm}
 \caption{MMSE periodograms for the signals detected in the GJ876 radial velocities.  From top to bottom we show the 
raw velocities, followed by the MMSE periodograms after subtracting each signal sequentially (61, 30, 15, 1.9, 10, and 124~days, 
respectively).}
\label{gj876_mmse}
\end{figure}

In Fig.~\ref{gj876_mmse} we show our MMSE analysis for GJ876 data from Keck (\citealp{rivera10}) and HARPS (\citealp{bonfils13}), 
where the HARPS data have been processed using the TERRA pipeline (\citealp{anglada-escude12a}), and in addition to the currently known 
signals originating from four planets, we highlight two new planet-like signals in the data.  In the top and second top panels we can see the 
strongest two signals, originating from the two gas giants known to be in resonance, with periods of around 61 and 30 days 
respectively.  The first of the new signals (planets?) has an orbital period of 15~days (third panel down in the figure), placing it in the 
region of stability found by \citeauthor{correia10}, but the mass of this planet would be 0.1~\mj, much larger than that speculated by 
\citeauthor{correia10} or \citeauthor{rivera10}.  Indeed, it was not possible to reach the $\chi^2$ values or rms values quoted in these 
works without including this signal in the data by following the standard radial velocity planet detection procedure of analysing 
a periodogram, selecting the strongest statistically significant signal, fitting that signal with a Keplerian, subtracting off that 
signal, and then reanalysing the residuals to hunt for addition frequencies.  

The omission of discussion of this signal from previous works is puzzling, since the signal has a semi-amplitude of 20~\ms.  Since 
this signal is circular, and if 
this signal were to arise from a genuine gas giant planet found to be in an island of stability due to the Laplace resonance, it would also 
indicate that the other planetary orbits could be more circular than previously thought.  In fact, we found the best fit by fixing all the orbits in the 
system to zero throughout the process.  The inclusion of this signal does not decrease the amplitudes of the previous planetary signals, 
meaning the masses previously found remain, only the orbits are circularised for some of them.  \citet{anglada-escude10} and \citet{wittenmyer12} 
have shown that resonant planets can be mistaken for less planets in a system but with higher orbital eccentricities.

Even though the MMSE spike for this 15~day signal is very significant, having a FAP lower than 10$^{-4}$, we still must ensure that it 
does not originate from any interference by the window function, especially since it is near half the lunar cycle period.  To test this 
we generated a fake system including only the first two planets, GJ876$b$ and $c$, which have the strongest signals prior to the 
15~day signal, fit them out and then see what remains in the data.  We keep the observed timestamps and uncertainties 
in this process.  We found that no significant signal at any frequency is present in the data after fitting out the first two signals.  
This is also the case if we inject all four of the previously known signals and rerun this experiment.  This tells us that the 15~day 
signal is present in the observed data set.

After considering the 15~day signal, the next strongest peak is the 1.93~day planet that was previously known (fourth panel down in 
Fig~\ref{gj876_mmse}), found to be significant in our combined data set.  However, after considering this planetary orbit, 
the next signal that emerges from the data is not the outer planet beyond 120 days, but a 10~day signal that we show in 
fifth panel of the figure.  We note that this signal was almost as significant as the 1.93~day signal after removal of the 
15~day signal from the data.  This signal again has a FAP of less than 10$^{-4}$, meaning it is statistically significant in the data.  
This signal has a semi-amplitude of 4.97~\ms, and if it was from an additional orbiting planet in this system, it would relate 
to a planet with a minimum mass of 0.02~\mj.

\begin{figure}
\vspace{5.5cm}
\hspace{-4.0cm}
\includegraphics{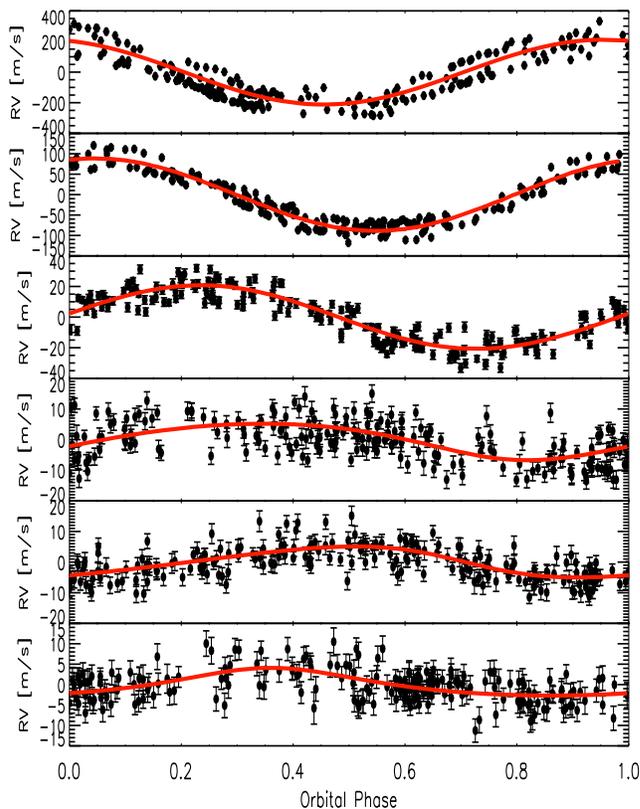}
\vspace{5.6cm}
 \caption{Phased radial velocity curves for all six signals detected in the combined Keck and HARPS-TERRA 
Doppler velocities.  From top to bottom we show signals with periods of 61, 30, 15, 1.92, 10, and 124~days, 
respectively.  The red curve shows the best fit Keplerian solutions we find for each of these signals.}
\label{gj876_phased}
\end{figure}

Finally in the sixth panel of Fig~\ref{gj876_mmse} we can see the signal from the outer planet with a period of 124.7~days.  This 
signal is well above the level of significance threshold to claim the existence of a signal, confirming 
this planet does still fit within the configuration we have discovered in the current data.  If we make the bold assumption 
that all of these signals are from orbiting planets in the GJ876 system, then the system contains a likely 8:4:2:1 Laplace resonance between 
four planets, three of which are gas giants orbiting this M-dwarf, something that is known to be rare, and two low-mass planets 
interior to the resonance, all of which have close to circular orbits.  However, given the 
total mass that the three largest planets would maintain (periods of 61, 30, and 15~days respectively), it is necessary to perform 
further dynamical simulations to test if such a system could be stable over the long term.  The previous works 
that do find islands of stability around 15~days do give us hope that at least one of these two signals originates from a planetary 
orbit, and if it is at 15~days, it is highly likely that we have found a new Laplace resonant chain.  The phase 
folded velocity curves for all six signals we have detected are shown in Fig.~\ref{gj876_phased}.

A non-planetary origin might be expected for the 15~day and 10~day signals since both the 
\citet{correia10} and \citet{rivera10} analyses failed to spot them when applying a Newtonian integration 
analysis that considers the dynamical interactions between the planets.  Although both these works suggested there 
are islands of stability around 15~days, \citet{gerlach12} suggest that these islands are likely not long-term stable given 
the planetary system configuration suggested by these works.  
However, they show that planets on orbits exterior to the currently known outer planet in this system could exist in 
long-term stable orbits.  We note that after all the previous signals have been considered we see two emerging 
spikes in the MMSE, one with a period of around 400~days and the other with a period of 1250~days.  These could be the 
first indications of additional longer-period planets in the GJ876 system, motivating continued observation of this 
highly prized and nearby planetary system. 

\begin{figure}
\vspace{5.5cm}
\hspace{-4.0cm}
\includegraphics{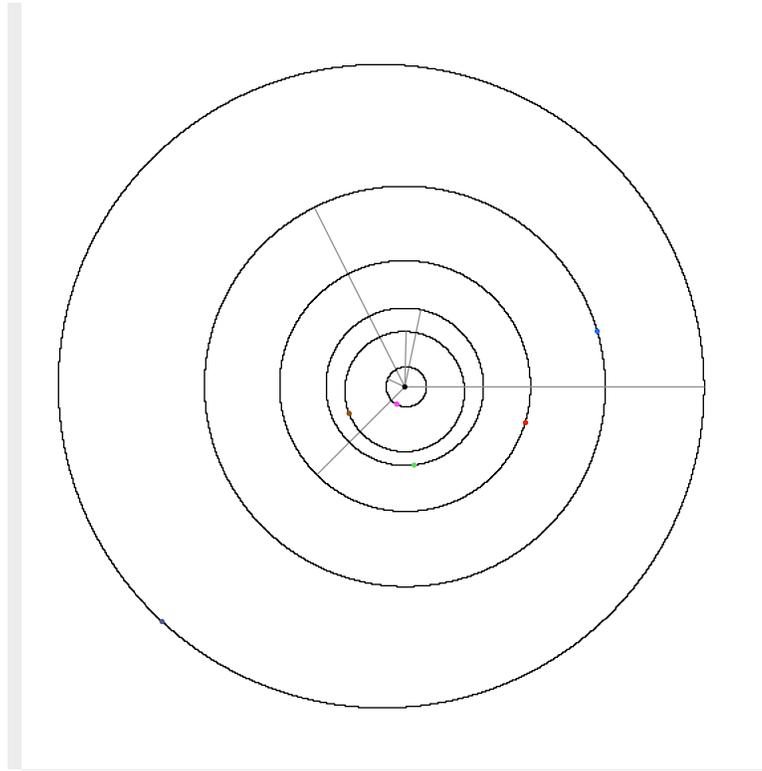}
\vspace{4.6cm}
 \caption{Orbital configuration of the possible planetary system orbiting GJ876.}
\label{orbits}
\end{figure}

There is the possibility that these signals are due to rotationally 
modulated spots on the stellar surface manifesting as an apparent velocity shift due to deformations of the spectral 
lines used to calculate the Doppler velocities.  Such activity induced signals can be measured and tracked using specific indicators like 
the Calcium HK chromospheric lines (\citealp{jenkins11}), or possibly even corrected for using measurements of the line bisectors 
(see \citealp{jenkins09}).  In the case of GJ876, both \citet{rivera05} and \citet{rivera10} show that the rotation period for this star 
is around 90-100 days, using photometry and Ca\sc ii \rm HK measurements respectively.  Therefore, it is unlikely that the new strong 
frequencies we see in the combined Keck and HARPS data are due rotational modulation of star spots, since we would 
be detecting harmonic frequencies beyond the 3rd-order.  
Finally, the Keck and HARPS-TERRA radial velocities are shown in Table~\ref{table:rvs}, 
the final system configuration is shown in Fig.~\ref{orbits} (output from the Systemic Console; \citealp{meschiari09}), assuming all originate from planetary orbits, 
and a list of the signal parameters are shown in Table~\ref{tab:system}.  All uncertainties were calculated using the 
Markov Chain Monte Carlo routines within the Systemic Console.

\subsection{Amplitude, Phase, and Number of Components}

In the previous sections we have described our procedure for signal detection in unevenly sampled timeseries that 
contains stationary and repeating signals, such as a radial velocity timeseries containing Doppler signals induced by planets 
orbiting a star.  We have also shown that the proposed MMSE method works at least as well as the classical LSP method 
of signal detection, and when applied to signal rich data like that for GJ876, it has no problem in detecting multiple 
signals in the data.  However, now we will show the power of the MMSE over the LSP method, in particular focusing on the 
additional information provided by our method.

Looking back at Eq$^{\rm{n}}$~\ref{eq:mmse}, we see that our MMSE method does not only focus on properties inherent in 
the spherical harmonics, or orthogonal sine and cosine functions, but it also minimises properties of the signal that 
are present in the unevenly sampled data set.  In particular, this method constrains the amplitude ($a$), the phase ($\phi$), 
and also the number of components that best describe the timeseries in the presence of noise ($N_C$).  Such information 
can be invaluable in timeseries analysis.  For example, when trying to confirm the nature of low-amplitude signals in a radial 
velocity timeseries one can assess the phase and amplitude as a function of time to ensure that these properties of the 
signal are more or less constant, which would be expected from a quasi-stationary source like 
stellar activity (see \citealp{dawson10}).  The number of components that best describe the data can also be used in this 
sense to estimate the upper-limit of planetary signals one should search for in the data and anything else can be considered as noise.

\subsection{Phase Test for GJ876$b$ and $c$}

Lets take the case of the confirmed gas giant planets orbiting GJ876.  The hypothesis is that if the two signals are more or less stationary, 
as they should be if they originate from the Doppler 
effect induced by the gravitational tug of these planets, then the phase offset ($O = \phi_1 - \phi_2$) between 
the two signals, or indeed the phase of each signal independently, should also be essentially stationary as a function of time, 
i.e. both signals conserve their phase across the baseline of the timeseries.  Therefore, with enough data across a time 
baseline long enough to confidently sample a double planet signal, we can measure the phase of both signals in the 
first half of the data set, do the same for the second half of the data set, and ensure that they are equal (i.e. 
$O_1 - O_2 = 0$), within some tolerance level, essentially ensuring that the phase does not vary with time.  A similar 
analysis can be performed on the amplitude of these signals, or of course, the frequencies aswell.  We have shown previously 
the significance of the signals for planets $b$ and $c$ orbiting GJ876 are very strong and given our current data we can 
clearly detect both of these signals in the first and second halves of the velocity timeseries independently.

\begin{figure}
\vspace{5.5cm}
\hspace{-4.0cm}
\includegraphics{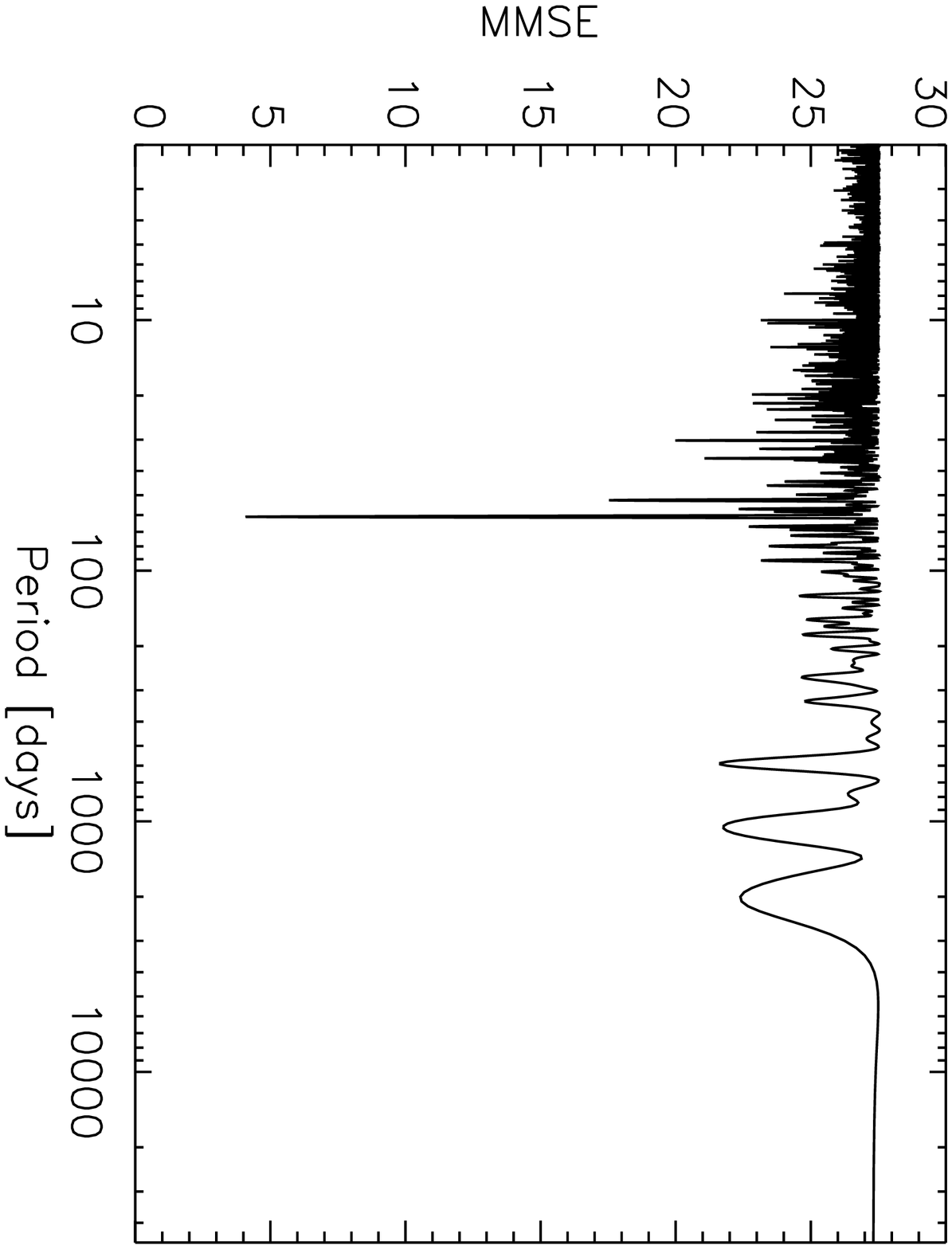}
\includegraphics{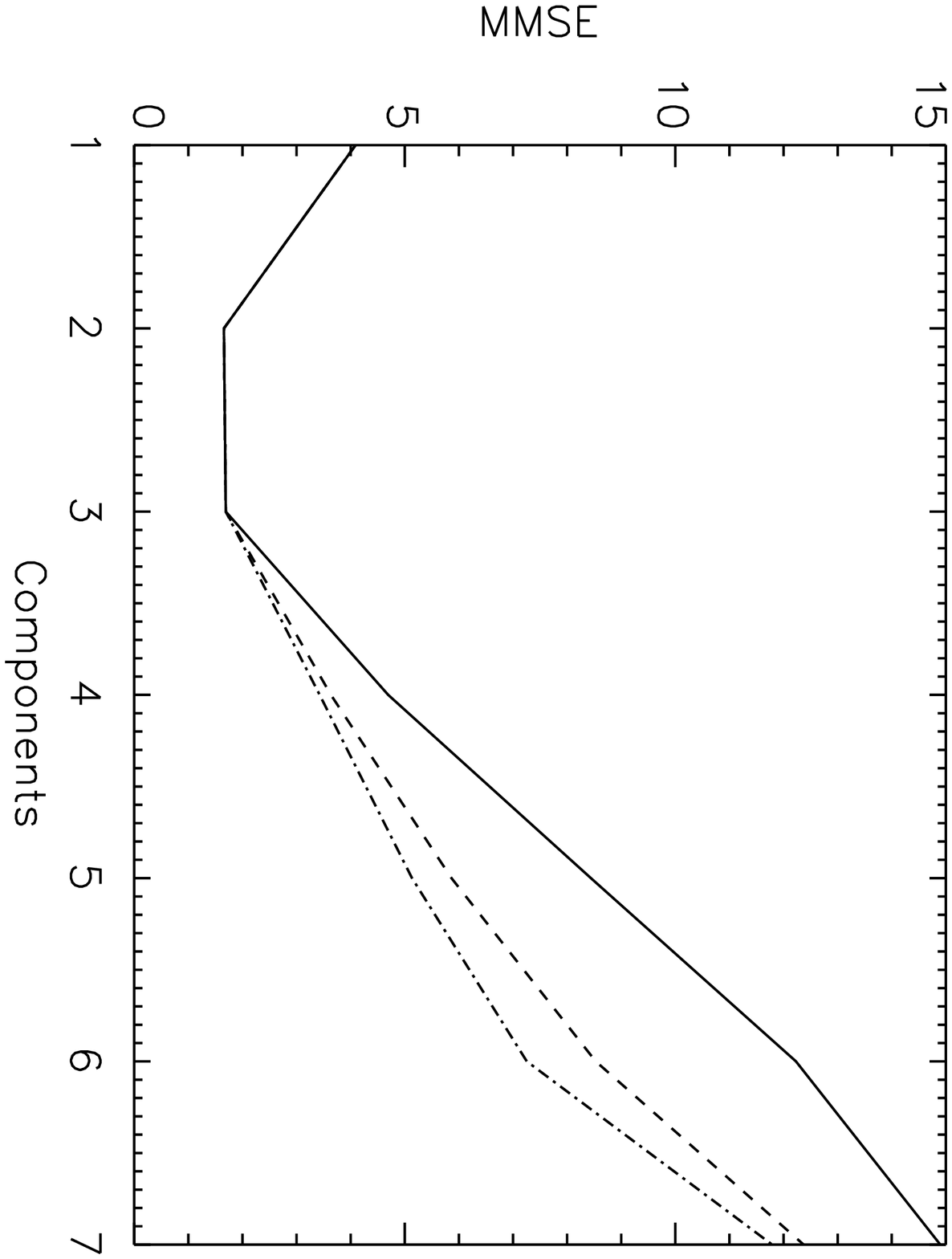}
\vspace{5.6cm}
 \caption{The top panel shows the MMSE periodogram for the full GJ876 data set from Keck.  The lower panel shows the 
results from the Trellis component analysis.  The three neighbourhood widths were 11 (solid curve), 13 (dashed curve), and 
15 (dot-dashed curve), meaning within the trellis analysis we chose to include 11, 13, and 15 neighbouring spikes on either 
side of the frequency spike we are testing.}
\label{gj876_mmse1}
\end{figure}

First of all, we ran the analysis as we did before on this data using the procedural steps we outlined in $\S$~3.  
For clarity we again show the MMSE periodogram in the top panel of Fig.~\ref{gj876_mmse1} for the full radial velocity data set, and we found both of the planetary frequency spikes to 
be in the first three strongest spikes, as expected.  We then selected a few of the deepest spikes and ran the component 
analysis, as we did before on the simulated data in $\S$~4.  Since the peaks we selected contain the two gas giant planetary signals, 
the component search returns a value of 
two, possibly three, components in the curves (lower panel in Fig.~\ref{gj876_mmse1}), for all three neighbourhood widths we chose.  This result not only adds weight to 
the reality of these signals, but also gives us confidence in our analysis procedure, since the component search rules out all other 
peaks we selected as being simply noise related.

Next, we proceeded to split the data into two chunks of equal length as a function of time and run the analysis again, clearly detecting 
both peaks in both chunks of data.  Again we get the same results from the component search, showing this method is very efficient at 
sifting through frequencies that are noise related and those that are genuine.  We then measured the phases of both signals from both 
chunks of data, essentially giving four phase measurements, two from the two signals in the first half of the timeseries and two from the 
two signals in the second half of the timeseries, and compared the difference of these two phases to test if the phase difference between 
the two signals is stationary as a function of time.  

\begin{figure}
\vspace{5.5cm}
\hspace{-4.0cm}
\includegraphics{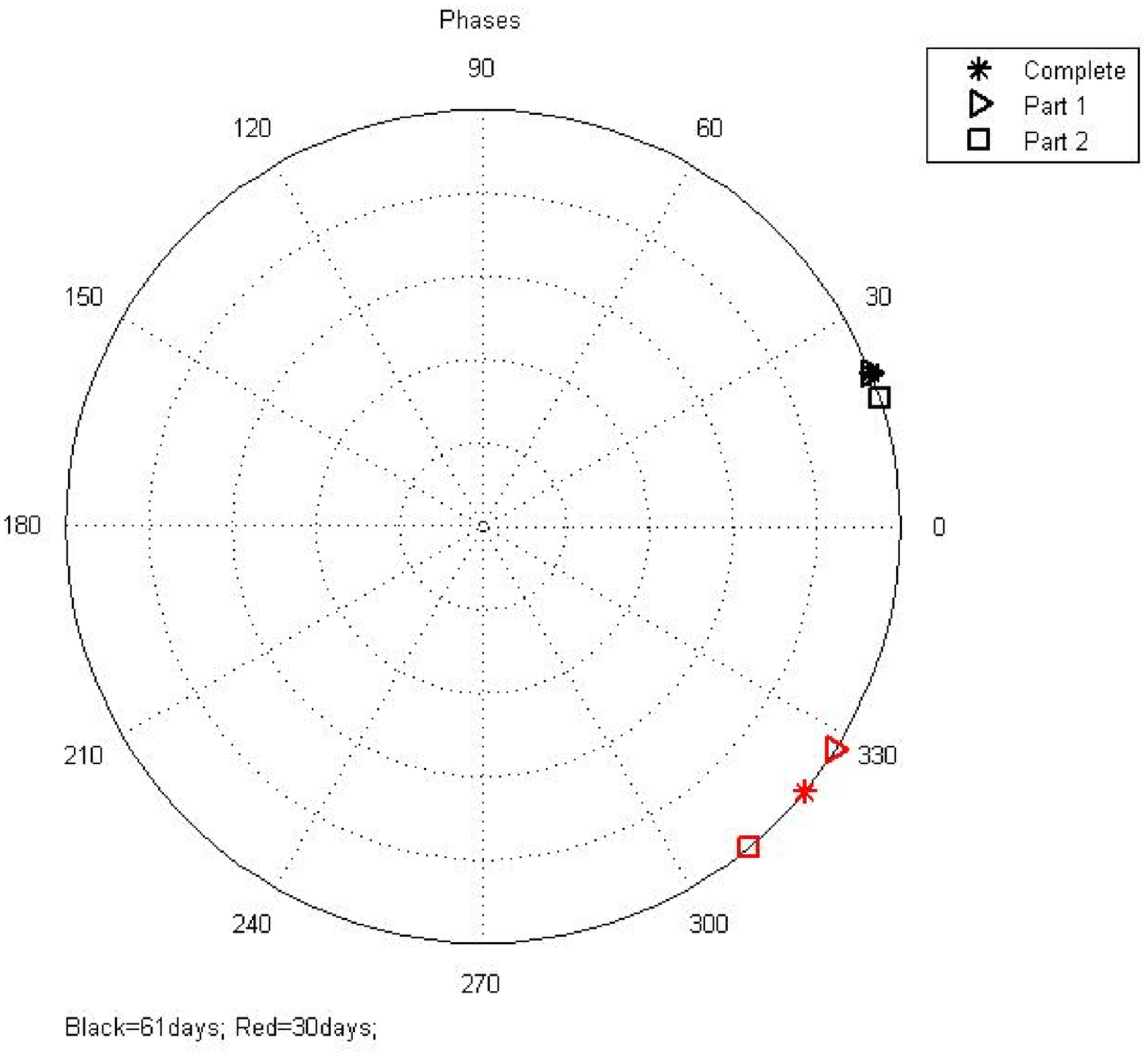}
\includegraphics{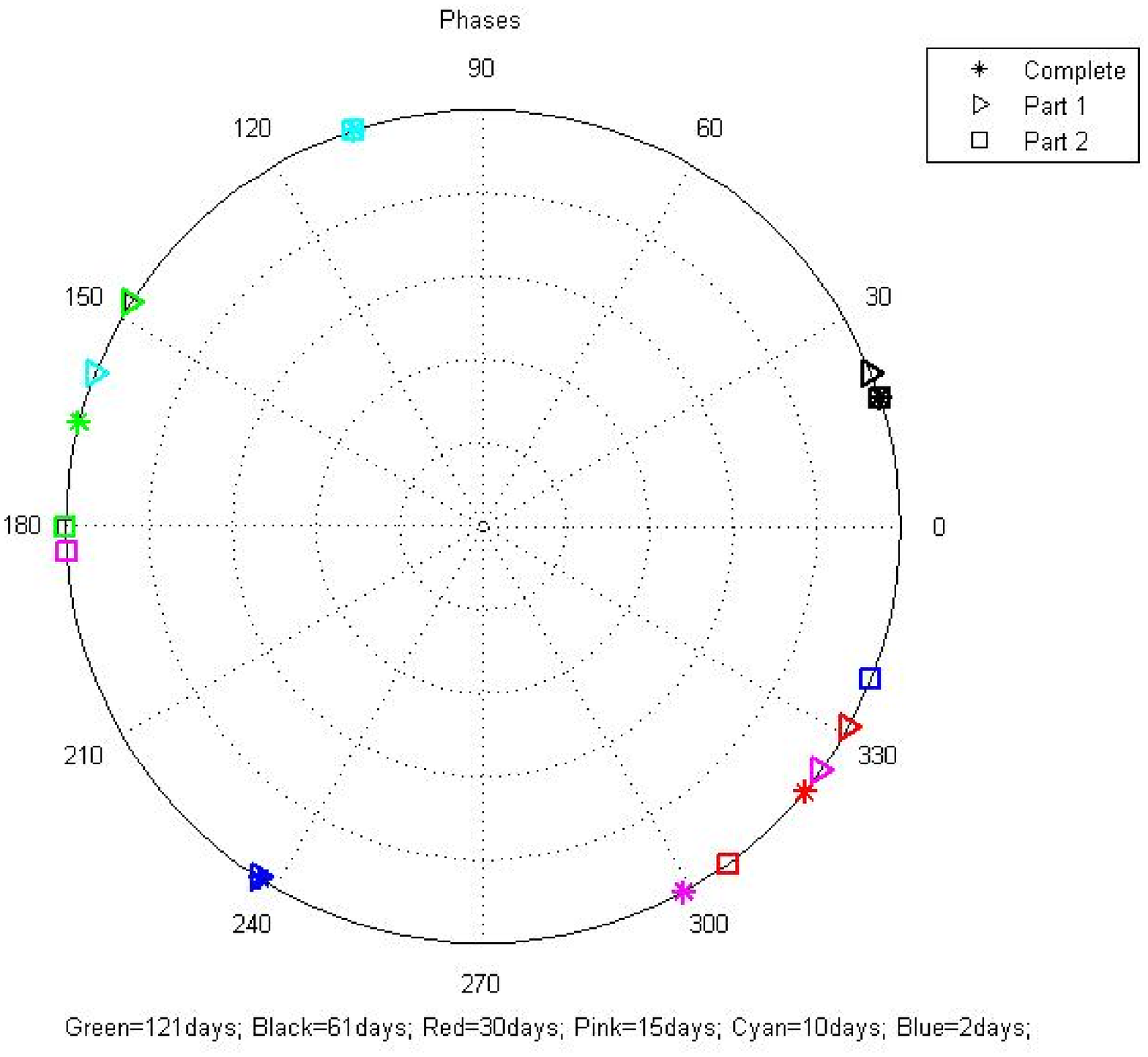}
\vspace{14.5cm}
 \caption{Phase tests for GJ876$b$ and GJ876$c$ are shown in black and red symbols respectively in the top plot, and in the bottom 
plot for all detected signals.  The symbols represent the measured phase of the detected MMSE signals 
for the complete timeseries (asterisks), the first half of the data set (open triangle), and the second half of the data set (open 
square).  The 360$^{o}$ phase angles are represented on a circular phase plot, with the symbols plotted at the extremum of the 
plot.  A key is also included in the top right that explains the symbols.  The colours of the symbols for all signals in the bottom panel 
are labeled as follows: 61 days (Black), 30 days (Red), 15 days (Pink), 2 days (Blue), 10 days (Cyan), and 121 days (Green)}
\label{phase_tests}
\end{figure}

In the top panel in Fig.~\ref{phase_tests} we show the results of these tests on a phase diagram and we can see that the symbols representing 
all the timeseries and only the first and second halves of the timeseries all have similar phase values, and hence the phase offset between both the planets 
does not change as a function of time.  This adds weight to the reality of the planetary system, showing that the signals are stable 
over the time baseline of the data set, expected from a true Doppler signal induced by an orbiting companion.  Therefore, our method 
allows secondary tests that can be used to confirm the reality of long-baseline data, and a similar analysis can be performed using the amplitudes 
of the signals, as these should also be stationary with time for a real Doppler signal.  A caveat to the method is that for signals that 
appear weak in the MMSE analysis, which in this case is the 30 day signal in the two halves of the data, the frequency sampling must 
be increased to ensure one selects the proper frequency peak to test.  This is because in the local neighbourhood of frequency peaks 
surrounding a genuine signal peak, the frequencies and amplitudes vary very little but the phase varies a lot, and hence one must 
be sure to select the correct peak originating from the genuine Doppler signal.

\begin{table*}
\center
\caption{Keck and HARPS-TERRA radial velocities for GJ876.}
\label{table:rvs}
\begin{tabular}{ccc}
\hline
\multicolumn{1}{c}{JD}& \multicolumn{1}{c}{RV} & \multicolumn{1}{c}{Error} \\
\multicolumn{1}{c}{days}& \multicolumn{1}{c}{m/s} & \multicolumn{1}{c}{m/s} \\ \hline
\underline{Keck} &  & \\
602.09311 &  329.19 &      2.68  \\
603.10836 &  345.30 &      2.70  \\
604.11807 &  335.99 &      2.77  \\
605.11010 &  336.00 &      2.78  \\
606.11129 &  313.94 &      2.75  \\
607.08450 &  288.73 &      2.75  \\
609.11637 &  197.05 &      2.81  \\
666.05032 &  338.38 &      2.68  \\
690.00713 &  -115.22 &      2.87  \\
715.96471 &  197.16 &      2.63  \\
785.70436 &  365.44 &      3.39  \\
983.04582 &  -65.63 &      2.74  \\
984.09389 &  -84.80 &      2.83 \\
1010.04457 &  -43.58 &      2.60  \\
1011.10207 &  -26.76 &      2.38  \\
1011.98546 &  -0.63 &      2.28  \\
1013.08905 &  26.92 &      2.70  \\
1013.96558 &  50.29 &      2.41  \\
1043.02045 &  -45.34 &      2.66  \\
1044.00022 &  -70.50 &      2.67  \\
1050.92784 &  -115.86 &      2.72  \\
1052.00302 &  -98.84 &      2.89  \\
1068.87655 &  -90.26 &      2.75  \\
1069.98405 &  -61.21 &      2.79  \\
1070.96594 &  -56.96 &      2.81  \\
1071.87782 &  -27.61 &      2.73  \\
1072.93848 &  -16.71 &      2.70  \\
1170.70376 &  -86.04 &      3.12  \\
1171.69171 &  -99.71 &      2.89  \\
1172.70252 &  -81.15 &      2.84  \\
1173.70148 &  -78.61 &      2.94  \\
1312.12727 &  -103.43 &      2.69  \\
1313.11723 &  -101.40 &      2.69  \\
1343.04074 &  68.57 &      2.80  \\
1368.00106 &  -147.61 &      2.84  \\
1369.00183 &  -156.59 &      2.89  \\
1370.05951 &  -139.31 &      2.61  \\
1372.05858 &  -120.61 &      3.54  \\
1409.98670 &  -49.88 &      2.75  \\
1410.94861 &  -48.68 &      2.81  \\
1411.92171 &  -56.68 &      2.66  \\
1438.80200 &  -23.37 &      2.84  \\
1543.70165 &  -105.40 &      2.96  \\
1550.70152 &  -161.29 &      2.77  \\
1704.10273 &  157.72 &      2.68  \\
1706.10773 &  107.70 &      2.86  \\
1755.98025 &  307.51 &      3.26  \\
1757.03786 &  281.33 &      2.94  \\
1792.82213 &  -179.43 &      2.79  \\
1883.72512 &  226.87 &      2.74  \\
1897.68199 &  88.28 &      2.83  \\
1898.70648 &  80.85 &      2.82  \\
1899.72426 &  70.86 &      2.77  \\
1900.70359 &  51.60 &      2.65  \\
2063.09867 &  245.77 &      2.93  \\
2095.02441 &  -194.59 &      2.95  \\
2098.05057 &  -230.27 &      3.09  \\
2099.09480 &  -222.44 &      3.08  \\
2100.06624 &  -234.09 &      2.70  \\
2101.99145 &  -214.27 &      2.77  \\
2128.91479 &  167.27 &      3.11  \\
2133.01847 &  92.78   &    3.00  \\
2133.88182 &  105.82 &      3.16  \\
\hline
\end{tabular}
\end{table*}

\begin{table*}
\begin{tabular}{ccc}
\hline
2160.89624 &  -226.42 &      2.81  \\
2161.86235 &  -231.15 &      3.03  \\
2162.88042 &  -195.02 &      3.11  \\
2188.90903 &  156.83    &   2.84  \\
2189.80815 &  152.59    &   3.16  \\
2236.69389 &  225.13    &   2.81  \\
2238.69635 &  248.60    &   2.92  \\
2242.71316 &  264.24    &   3.16  \\
2446.07064 &  117.43    &   3.19  \\
2486.91681 &  230.06    &   2.54  \\
2487.12395 &  218.81    &   2.47  \\
2487.91836 &  216.56    &   2.60  \\
2488.12695 &  222.24    &   2.48  \\
2488.94388 &  199.66    &   2.19  \\
2514.86655 &  -84.38    &   3.19  \\
2515.87295 &  -108.41  &     3.00  \\
2535.77402 &  83.48      & 3.05  \\
2536.02384 &  87.44      & 2.87  \\
2536.80403 &  114.45    &   3.03  \\
2537.01292 &  117.70   &    2.76  \\
2537.81194 &  126.07   &    2.80  \\
2538.01383 &  133.42   &    3.00  \\
2538.80067 &  162.08   &    3.07  \\
2539.92124 &  175.89   &    2.79  \\
2572.71250 &  -7.07     &  2.67  \\
2572.91949 &  -15.40   &    2.77  \\
2573.74296 &  -28.24   &    2.65  \\
2573.87844 &  -28.58   &    2.62  \\
2574.76393 &  -66.20   &    2.61  \\
2574.94041 &  -64.63   &    2.67  \\
2575.71905 &  -87.17   &    2.56  \\
2600.75110 &  172.51   &    2.42  \\
2601.75071 &  177.92   &    2.48  \\
2602.72004 &  201.51   &    2.52  \\
2651.71796 &  -78.56   &    3.91  \\
2807.02778 &  202.50   &    2.79  \\
2829.00786 &  -205.85 &      2.81  \\
2832.07993 &  -137.56 &      2.95  \\
2833.96325 &  -87.32   &    2.91  \\
2835.08493 &  -51.90   &    2.74  \\
2848.99946 &  188.48   &    3.25  \\
2850.00090 &  171.94   &    3.01  \\
2851.05668 &  166.92   &    3.10  \\
2854.00713 &  130.49   &    2.92  \\
2856.01579 &  155.93   &    3.00  \\
2897.82598 &  -2.50     &  2.87  \\
2898.81468 &  29.81     &  2.88  \\
2924.79475 &  255.62   &    3.03  \\
2987.71613 &  246.15   &    3.58  \\
2988.72395 &  239.18   &    3.07  \\
3154.11700 &  95.22     &  3.00  \\
3181.00540 &  -14.91   &    3.00  \\
3181.11618 &  -22.60   &    2.90  \\
3182.06964 &  -60.62   &    2.79  \\
3191.03662 &  -220.60 &      2.94  \\
3195.97023 &  -140.86 &      2.94  \\
3196.99712 &  -122.80 &      3.11  \\
3301.83459 &  13.71     &  2.37  \\
3302.72877 &  -36.34   &    2.31  \\
3303.78576 &  -85.95   &    2.36  \\
3338.74360 &  119.33   &    2.91  \\
3367.71753 &  -180.10 &      3.12  \\
3368.71947 &  -193.13 &      3.31  \\
3369.70562 &  -192.62 &      2.54  \\
3547.08770 &  -98.83   &    2.64  \\
3550.10132 &  -148.44 &      2.99  \\
3551.10883 &  -166.49 &      2.94  \\
\hline
\end{tabular}
\end{table*}

\begin{table*}
\begin{tabular}{ccc}
\hline
3552.09458 &  -165.40 &      2.89  \\
3571.00432 &  27.08     &  2.86  \\
3603.07412 &  104.97   &    3.19  \\
3604.01707 &  51.89     &  2.93  \\
3604.98440 &  13.84     &  3.16  \\
3724.71853 &  88.50     &  2.80  \\
4083.77192 &  287.58   &    3.28  \\
4337.09682 &  -0.28     &  2.99  \\
4343.91291 &  -109.95 &      3.10  \\
4345.11351 &  -93.51   &    2.96  \\
4396.78303 &  27.97     &  2.88  \\
4397.84797 &  -4.46     &  2.99  \\
4634.10109 &  182.38   &    2.94  \\
4636.08653 &  132.28   &    2.91  \\
4639.07368 &  64.12     &  2.79  \\
4667.05860 &  -137.50 &      2.8  \\
4672.99139 &  -31.83   &    2.54  \\
4675.95043 &  73.06     &  2.45  \\
4687.00220 &  318.88   &    2.80  \\
4702.07311 &  9.81       & 2.64  \\
4703.06794 &  -0.30     &  2.29  \\
4704.02881 &  -25.07   &    2.53  \\
4819.80176 &  99.74     &  2.68  \\
4822.78895 &  25.41     &  2.51  \\
5051.08321 &  284.20   &    2.56  \\
5054.03189 &  274.77   &    3.01  \\
5143.87496 &  -127.21 &      2.39  \\
5166.83669 &  250.15   &    2.41  \\
5168.79585 &  272.52   &    2.58  \\
5200.76205 &  -59.18   &    2.52  \\
5201.76137 &  -78.29   &    2.42  \\
5202.73775 &  -103.93 &      2.28  \\
\\ \hline
\underline{HARPS} & & \\
          3339.55769  &    210.90  &      2.18  \\
          3542.76959  &    175.44  &      2.19  \\
          3542.93864  &    162.55  &      2.10  \\
          3543.75050  &    112.82  &      2.09  \\
          3543.87080  &    108.38  &      2.07  \\
          3543.94632  &    106.84  &      2.09  \\
          3544.74647  &     72.35   &     2.17  \\
          3544.87437  &     67.05   &     2.06  \\
          3544.94455  &     63.42   &     2.08  \\
          3545.78504  &     28.25   &     2.66  \\
          3546.73885  &      0.00    &    2.58  \\
          3546.88000  &     -7.65   &     2.06  \\
          3546.94738  &    -10.10  &      2.14  \\
          3547.73491  &    -33.05  &      2.32  \\
          3547.86712  &    -35.70  &      2.07  \\
          3547.94555  &    -34.82  &      2.13  \\
          3550.73512  &    -76.60  &      2.18  \\
          3550.84190  &    -79.72  &      2.08  \\
          3550.94392  &    -83.54  &      2.11  \\
          3572.85339  &     89.75   &     2.25  \\
          3573.77291  &     87.62   &     2.09  \\
          3576.74516  &     87.28   &     2.06  \\
          3577.71674  &    109.28  &      2.06  \\
          3577.89513  &    109.51  &      2.07  \\
          3579.86570  &    149.02  &      2.06  \\
          3669.66480  &    -33.22  &      2.10  \\
          3670.50887  &    -38.22  &      2.08  \\
          3670.71111  &    -43.30  &      2.09  \\
          3671.50963  &    -63.62  &      2.07  \\
          3671.71358  &    -63.10  &      2.12  \\
          3672.51327  &    -64.64  &      2.05  \\
          3672.70490  &    -70.00  &      2.07  \\
\hline
\end{tabular}
\end{table*}

\begin{table*}
\begin{tabular}{ccc}
\hline
          3673.51828  &    -84.75  &      2.10  \\
          3673.70526  &    -83.97  &      2.07  \\
          3674.50616  &    -81.30  &      2.05  \\
          3674.69603  &    -87.32  &      2.10  \\
          3675.51016  &    -95.90  &      2.09  \\
          3975.72341  &    -29.59  &      2.06  \\
          3979.81110  &    -64.72  &      2.19  \\
          4228.89045  &     22.46   &     2.09  \\
          4260.93227  &    463.58  &      2.35  \\
          4291.86547  &     29.54   &     2.13  \\
          4295.79232  &    -25.12  &      2.04  \\
          4298.80154  &    -44.60  &      2.09  \\
          4339.71897  &     18.95   &     2.07  \\
          4344.75827  &    -14.39  &      2.07  \\
          4392.62420  &    206.68  &      2.08  \\
          4422.54302  &    -40.55  &      2.14  \\
          4429.56649  &     66.48   &     2.11  \\
          4446.56696  &    372.04  &      2.08  \\
          4704.71839  &     48.06   &     2.08  \\
          4770.68394  &     94.21   &     2.14  \\
\end{tabular}
\end{table*}

\begin{table*}
\center
\caption{Keplerian solutions for GJ876.}
\label{tab:system}
\begin{tabular}{ccccccc}
\hline
\multicolumn{1}{c}{Parameter}& \multicolumn{1}{c}{GJ876~$b$} & \multicolumn{1}{c}{GJ876~$c$}   & \multicolumn{1}{c}{GJ876~$d^*$} & \multicolumn{1}{c}{GJ876~$e$} &
\multicolumn{1}{c}{GJ876~$f^*$}& \multicolumn{1}{c}{GJ876~$g$} \\ \hline
\hline

$P$ [days] & 61.03$\pm$3.81 & 30.23$\pm$0.19 & 15.04$\pm$0.04 & 1.94$\pm$0.001 & 10.01$\pm$0.02 & 124.69$\pm$90.04  \\
$e$ & 0.000$\pm$1x10$^{-3}$ & 0.002$\pm$1x10$^{-3}$ & 0.007$\pm$0.004 & 0.081$\pm$0.040 & 0.090$\pm$0.046 & 0.073$\pm$0.048  \\
$\omega$ [$^o$] & 116.7$\pm$1.1 & 225.2$\pm$5.0 & 78.3$\pm$10.3 & 157.4$\pm$23.3 & 88.0$\pm$10.5 & 360.0$\pm$15.8 \\
M$_{0}$ [$^o$] & 259.5$\pm$1.1 & 117.9$\pm$5.0 & 198.5$\pm$10.4 & 79.9$\pm$23.7 & 108.3$\pm$10.5 & 230.2$\pm$16.3 \\
$K$ [ms$^{-1}$] & 211.57$\pm$32.92 & 88.72$\pm$13.81 & 20.71$\pm$3.24 & 5.91$\pm$0.98 & 5.00$\pm$0.80   &   3.37$\pm$0.53 \\
$m_{p} \sin i$ [M$_{\rm{J}}$] & 1.927$\pm$0.003 & 0.637$\pm$0.002 & 0.118$\pm$0.002 & 0.017$\pm$0.001 & 0.025$\pm$0.001 & 0.039$\pm$0.001 \\
\hline
$\gamma_{\rm{Keck}}$ [ms$^{-1}$] &  -49.68$\pm$0.50 & &&&& \\
$\gamma_{\rm{HARPS-TERRA}}$ [ms$^{-1}$] &  -79.23$\pm$0.63 & &&&& \\
$\sigma_{J}$ [ms$^{-1}$] & 1.98$\pm$ \\
rms [ms$^{-1}$] & 3.33 \\
$\chi^2$ & 1.787 \\

\hline
\end{tabular}
\caption{$^*$ Assumes the signal relates to a Doppler shift induced by an orbiting planet.}
\end{table*}

This test shows that our MMSE method allows us to confirm the nature of signals as 1) being real signals present in the data, and 
2) being signals that are stable in phase as a function of time, increasing the likelihood that they are real Doppler signals and not 
quasi-static signals originating from time-varying phenomena.  The caveat here is that the signals being tested must be significant 
in both halves of the timeseries so that one can select them unambiguously in the data.  This analysis can also be performed using 
the signal amplitudes and frequencies.  These features of our MMSE method present a step beyond single LSP-like 
methods, as addition information is extracted from the data in the processing of the periodogram.

Finally, since we believe the phase analysis can be used as a robust way to determine if signals are planets or not, we apply 
it to the rest of the sample to test if the other signals we detect conserve their phase.  The lower plot in Fig.~\ref{phase_tests} shows 
the phases calculated as before but including the additional four signals.  We note that the other signals are not as significant as the first 
two but we believe we can select the correct frequency spike from both halves of the data series.

The most significant of the additional four signals is the 15~day signal and as it is in the next resonant site for a real planet, there 
is an additional reason to believe it could be real.  We show the measured phases in pink in the figure and we find that the complete data 
and the first half data are in good agreement, however the second half of the data shows a phase that is around 120 degrees away 
from the complete data, throwing doubt on the nature of the signal.  However, it must be remembered that this would be the smallest of the 
three gas giant planets in the resonance.  

When we look at the other signals in this manner we note that the planet at $\sim$2~days shows some phase variations at a similar 
level to that of the 15~day signal.  Again, the amplitude of this signal is even lower than the previous ones so its significance is 
lower and makes it difficult to select the correct spikes in the analysis in half of the data.  The phase of the signal at 10~days shows some 
variation, albeit at a lower level than the 2 and 15~days signals, whereas the 124~day planetary signal clearly conserves the phase 
in both halves of the data, helping to confirm it's reality as a Doppler signal.

Although the 2, 15, and 10~day signals exhibit some phase variation with time, at least some of this could be due to the dynamical 
nature of the system, especially the Laplace resonance.  However, we also note that if we just include the three signals with periods 
of 61, 30, and 2~days, we find the phase to be more stable for the 2 day planetary signal.  Therefore, it could be that including signals 
that are not stable Doppler signals induces phase variations on other stable signals.  Also it could be the case that the 2 and 15~day 
signals are correlated at some level, meaning when we analyse them both in parallel the correlations manifest as phase variations 
in the timeseries.

\section[summary]{Summary and Future Work}

We have developed and tested a new method for detecting and analysing possible signals in unevenly sampled timeseries data, with 
a particular emphasis on the analysis of precision radial velocities to search for low-mass exoplanets.  Our method begins with a new 
periodogram like analysis of the data, using a minimum mean squared error approach that allows us to determine the amplitude and 
phase of the possible signal directly.  We then select the periodogram spikes we want to analyse and make use of a trellis-type analysis to test 
which of these peaks are related to real signals and which are noise spikes attributed to interference from the window function.  Our method 
is a global approach that hunts for all signals in the full data set, given by the distribution of frequency spikes measured in the data.  The trellis analysis 
is a generic solution that is only limited in scope by the available computing power.  
Finally, we can monitor the amplitudes and phases as a function of time to test if the signals we have found are stationary signals, like 
those from a Doppler source, or if they are non-stationary, arising from aliases or from external quasi-stationary sources like spot modulated 
activity on the star.

The following seven points summerise the benefits of our MMSE approach:

\begin{enumerate}

\item The MMSE method offers a unified framework to estimate the optimum number of components, and their parameters, in an unevenly sampled data set.

\item The MMSE approach attempts to estimate the global optimal solution (i.e. number of components, their frequencies, amplitudes, and phases) by using 
an analysis based on a simultaneous set of components and eliminating the dependence on the order in which the strongest frequencies are selected.

\item The model behind the MMSE method is more complete than standard periodogram approaches since it takes into consideration that the uneven sampling 
distortion depends on all the true components in the observed signal.

\item The MMSE approach is not an iterative process and no subjective stop criterion is required.  However, a gradient-based approach can be performed 
without considering the trellis component search part of the process, allowing a direct comparison of results gained from a LSP-like analysis.

\item No information about the window function is required.

\item The MMSE is robust against false-positives and false-negatives.

\item The MMSE method is also a powerful and generic tool to validate Doppler-like signals found by other methods.

\end{enumerate}

We applied our method to the Keck and HARPS data for the M-dwarf planet host star GJ876, known to host a system of planets that contains at least two short period gas giants.  
We studied this system because the large amplitude signals could be detected in both halves of the timeseries separately.  This allowed us to 
study the phase of the signal as a function of time, showing that the phase difference between both planets is stable over the length of the 
timeseries and therefore adding weight to the reality of these signals.  This analysis shows the power of our method over the LSP, since this gives no information 
on the signal parameters except for the frequency.

Further to this, we find an additional two planet-like signals in the data, having periods of 10 and 15~days, the latter of which could contain 
islands of stability linked by a 8:4:2:1 chain resonance.  However, given previous Newtonian integrational methods failed to spot these strong 
signals, we require further dynamical analyses to confirm if such a planetary configuration is stable over the long-term.  Indeed, we find 
phase variations with time for both of these signals, along with the signal at 2~days also, which could throw doubt on the origin of these signals 
as being from orbiting planets.  However since the signals are weak, this analysis is as yet inconclusive.

Finally, we plan to continue building our analysis algorithm, in particular working on a better understanding of the noise.  We aim to add a red noise 
model component that will allow us to better deal with correlations in the noise and between the data points, likely using a moving average model 
that has recently been shown to deal well with short term correlations between precision radial velocities (\citealp{tuomi13b}; \citealp{anglada-escude13}; 
\citealp{jenkins13b}).

\section*{Acknowledgments}

We thank our referee Artie Hatzes for a very thorough review of our work.  
We acknowledge useful discussion with Hugh Jones and Mikko Tuomi.  Jenkins acknowledges funding by Fondecyt through grant 3110004, partial support 
from Centro de Astrof\'\i sica FONDAP 15010003, the  GEMINI-CONICYT FUND, from the Comit\'e Mixto ESO-GOBIERNO DE CHILE, and from the 
Basal-CATA grant.  Yoma, Rojo, Mahu, and Wuth acknowledge the support of the Conicyt Team Research in Science and Technology Proyect ACT1120.  
Rojo also acknowledges support by Fondecyt through grant 1120299.

\bibliographystyle{mn2e}
\bibliography{refs}

\label{lastpage}

\end{document}